\documentclass[12pt,preprint]{aastex}
\usepackage{color}
\usepackage{amsmath}
\usepackage[symbol]{footmisc}

\def\deg{\ifmmode {^\circ}\else {$^\circ$}\fi}
\def\degree{\ifmmode {^\circ}\else {$^\circ$}\fi}
\def\mum{\ifmmode {\rm \,\mu {\rm m}}\else $\rm \,\mu {\rm m}$\fi}
\def\arcsec{\ifmmode ^{\prime \prime}\else $^{\prime \prime}$\fi}

\def\inch{\ifmmode ^{\prime \prime}\else $^{\prime \prime}$\fi}
\def\msunyr{\ifmmode {M_{\odot}~{\rm yr^{-1}}}\else $M_{\odot}~{\rm yr^{-1}}$\fi}
\def\msun{\ifmmode {M_{\odot}}\else $M_{\odot}$\fi}
\def\rsun{\ifmmode {R_{\odot}}\else $R_{\odot}$\fi}
\def\lsun{\ifmmode {L_{\odot}}\else $L_{\odot}$\fi}
\def\mstar{\ifmmode {M_{\star}}\else $M_{\star}$\fi}
\def\rstar{\ifmmode {R_{\star}}\else $R_{\star}$\fi}
\def\tstar{\ifmmode {T_{\star}}\else $T_{\star}$\fi}
\def\lstar{\ifmmode {L_{\star}}\else $L_{\star}$\fi}
\def\md{\ifmmode {M_d}\else $M_d$\fi}
\def\ld{\ifmmode {L_d}\else $L_d$\fi}
\def\ad{\ifmmode A_d\else $A_d$\fi}
\def\ldlstar{\ifmmode L_d / L_\star\else $L_d / L_{\star}$\fi}
\def\rearth{\ifmmode {\rm R_{\oplus}}\else $\rm R_{\oplus}$\fi}
\def\mearth{\ifmmode {\rm M_{\oplus}}\else $\rm M_{\oplus}$\fi}
\def\qdstar{\ifmmode Q_D^\star\else $Q_D^\star$\fi}
\def\kms{\ifmmode {\rm km~s^{-1}}\else $\rm km~s^{-1}$\fi}
\def\ms{\ifmmode {\rm m~s^{-1}}\else $\rm m~s^{-1}$\fi}
\def\mesc{\ifmmode m_{esc}\else $m_{esc}$\fi}
\def\rmin{\ifmmode r_{min}\else $r_{min}$\fi}
\def\rmax{\ifmmode r_{max}\else $r_{max}$\fi}
\def\mmin{\ifmmode m_{min}\else $m_{min}$\fi}
\def\mmax{\ifmmode m_{max}\else $m_{max}$\fi}
\def\rmind{\ifmmode r_{min,d}\else $r_{min,d}$\fi}
\def\rmaxd{\ifmmode r_{max,d}\else $r_{max,d}$\fi}
\def\mmaxd{\ifmmode m_{max,d}\else $m_{max,d}$\fi}
\def\vrad{\ifmmode v_{rad}\else $v_{rad}$\fi}
\def\qz{\ifmmode q_{0}\else $q_{0}$\fi}
\def\qi{\ifmmode q_{i}\else $q_{i}$\fi}
\def\ql{\ifmmode q_{l}\else $q_{l}$\fi}
\def\qs{\ifmmode q_{s}\else $q_{s}$\fi}
\def\rbrk{\ifmmode r_{brk}\else $r_{brk}$\fi}
\def\rdamp{\ifmmode r_{damp}\else $r_{damp}$\fi}
\def\rin{\ifmmode r_{in}\else $r_{in}$\fi}
\def\rout{\ifmmode r_{out}\else $r_{out}$\fi}
\def\tin{\ifmmode t_{in}\else $t_{in}$\fi}
\def\tout{\ifmmode t_{out}\else $t_{out}$\fi}
\def\ain{\ifmmode a_{in}\else $a_{in}$\fi}
\def\aout{\ifmmode a_{out}\else $a_{out}$\fi}
\def\r0{\ifmmode R_{0}\else $R_{0}$\fi}
\def\m0{\ifmmode m_{0}\else $m_{0}$\fi}
\def\M0{\ifmmode M_{0}\else $M_{0}$\fi}
\def\xm{\ifmmode x_{m}\else $x_{m}$\fi}
\def\sigz{\ifmmode \Sigma_0\else $\Sigma_0$\fi}
\def\gyr{\ifmmode {\rm g~yr^{-1}}\else ${\rm g~yr^{-1}}$\fi}
\def\cms{\ifmmode {\rm cm~s^{-1}}\else ${\rm cm~s^{-1}}$\fi}
\def\gcms{\ifmmode {\rm g~cm^{-2}}\else $\rm g~cm^{-2}$\fi}
\def\gcmss{\ifmmode {\rm g~cm^{-2}~s^{-1}}\else $\rm g~cm^{-2}~s^{-1}$\fi}
\def\gcmc{\ifmmode {\rm g~cm^{-3}}\else $\rm g~cm^{-3}$\fi}
\def\dcm2{\ifmmode {\rm dyn~cm^{-2}}\else $\rm dyn~cm^{-2}$\fi}
\def\ecsk{\ifmmode {\rm erg~cm^{-1}~s^{-1}~K^{-1}}\else $\rm erg~cm^{-1}~s^{-1}~K^{-1}$\fi}
\def\cm2{\ifmmode {\rm cm^{-2}}\else $\rm cm^{-2}$\fi}
\def\atilin{\ifmmode {\tilde{a}_{in}}\else $\tilde{a}_{in}$\fi}
\def\atilout{\ifmmode {\tilde{a}_{out}}\else $\tilde{a}_{out}$\fi}
\def\atil{\ifmmode {\tilde{a}}\else $\tilde{a}$\fi}
\def\ttil{\ifmmode {\tilde{t}}\else $\tilde{t}$\fi}
\def\sqrttt{\ifmmode {\tilde{t}^{1/2}}\else $\tilde{t}^{1/2}$\fi}

\def\gcc{g\,cm$^{-3}$}

\def\h2o{H$_2$O}
\def\sio2{SiO$_2$}
\def \ms{m\,s$^{-1}$\,}
\def \kms{km\,s$^{-1}$}
\def \msun{M$_{\odot}$}
\def \rsun{R$_{\odot}$}
\def \lsun{L$_{\odot}$}

\def \mearth{M$_{\oplus}~$}

\begin{document}
\shorttitle{Efficiency of Planetesimal Accretion}
\shortauthors{Podolak, Haghighipour, Bodenheimer, Helled, Podolak}

\title{Detailed Calculations of the Efficiency of Planetesimal 
Accretion in the Core-Accretion Model}
\author{Morris Podolak\altaffilmark{1}, Nader Haghighipour\altaffilmark{2}, Peter Bodenheimer\altaffilmark{3}, Ravit Helled\altaffilmark{4}, and Esther Podolak\altaffilmark{5}}

\altaffiltext{1}{Department of Geosciences, Tel Aviv University, Tel Aviv, Israel 69978}
\altaffiltext{2}{Institute for Astronomy, University of Hawaii-Manoa, Honolulu, HI 96822}
\altaffiltext{3}{UCO/Lick Observatory, University of California Santa Cruz, USA}
\altaffiltext{4}{Institute for Computational Science, Center for Theoretical Astrophysics and Cosmology, University of Z\"urich Z\"urich, Switzerland}
\altaffiltext{5}{Sysnet, Tel Aviv, Israel}

\begin{abstract}
We present results of a study of the accretion rate of planetesimals by a growing proto-Jupiter in the core-accretion model. 
The purpose of our study is to test the assumptions of Pollack et al. (1996) regarding the flux of planetesimals and their 
encounter velocities with the protoplanet. Using a newly developed code, we have accurately calculated planetesimals 
trajectories during their passage in the envelope by combining detailed three-body integrations with gas drag. To be 
consistent with Pollack et al., our calculations do not include the effect of nebular gas. Results point to several 
new findings. For instance, we find that only $4-5\,M_{\oplus}$ is accreted in the first 1.5 Myr before the onset of 
rapid gas accretion and $\sim 10\,{M_{\oplus}}$ is accreted simultaneously during this phase. We also find that mass 
accretion remains small ($0.3-0.4\,M_{\oplus}$) for $\sim 1$ Myr after this time. This late accretion, together with 
a rapid in-fall of gas could lead to a mixing of accreted material throughout the outer regions which  may explain 
the enhancement of high-Z material in Jupiter's envelope. Results demonstrate that encounters with the protoplanetary 
envelope become so fast that in most cases, ram pressure breaks up planetesimals. As a result, the accretion rate is 
largely independent of the planetesimals size and composition. We also carried out some calculations considering 
nebular gas drag. As expected, the accreted mass of planetesimals depended strongly on their size and composition. 
In general, nebular gas lowered the amount of accreted planetesimals, although the majority of planetesimals were 
still accreted during the rapid gas accretion phase.
\end{abstract}

\section{Introduction}
Planetesimal accretion is an integral part of planet formation.  Not only is it the fundamental mechanism for building the protoplanetary core in the core accretion scenario (CA), but it also contributes to the planet luminosity in the gas accretion phase.  The variation of the accretion rate during the period of protoplanet growth will determine both the composition of the envelope, and the time required to reach the collapse phase \citep{pollack96, iaros07, movsh10, d'angelo14, Venturini2016, lozovsky17}.  For that reason, determining the details of the planetesimal accretion rate is of fundamental importance.

The rate of planetesimal accretion can be broadly taken to consist of two processes.  The first is the trajectory of the planetesimal about the Sun.  This determines if, when, and at what speed the planetesimal will encounter the planetary envelope.  The second is the trajectory of the planetesimal through that envelope.  The outcome of an encounter between a planetesimal and the protoplanetary envelope is governed by the strength of the gas drag the planetesimal experiences.  This, in turn, depends on the planetesimal's velocity relative to the envelope, and the temperature and density of the gas it encounters.  In addition, if the gas drag heats the planetesimal, there will be mass lost due to ablation.  This not only leads to mass deposition in the envelope, it also increases the efficacy of gas drag in slowing the planetesimal.  A related capture mechanism must also be considered: If the stress due to the ram pressure on the planetesimal is higher than the mechanical strength of the material, the planetesimal can break up into smaller pieces which may then be easily stopped by the ambient gas.  Because the magnitude of the drag force acting on a planetesimal is directly related to the planetesimal's velocity relative to the gas and the density of that gas, its trajectory and velocity must be calculated accurately in order to correctly determine the outcome of its encounter with the gas envelope.

Each of these processes affects the other.  As mentioned above, the trajectory about the Sun determines the parameters of the encounter with the envelope.  These, in turn, determine the interaction of the planetesimal with the envelope, and whether it is captured or not.  Finally, even if the planetesimal escapes the envelope, it will have lost energy in the encounter with the envelope, and its subsequent trajectory about the Sun will be different. 

The interaction of a planetesimals with the envelope is driven by
the drag force of the envelope. This drag force depends on the size and material composition of the planetesimal, as well as the temperature of the envelope gas.  But, more importantly, it strongly depends on the velocity with which the planetesimal enters the envelope.  In the original work of \citet[hereafter called P96]{pollack96}, the rate of planetesimal encounters was calculated using a gravitational enhancement factor, $F_g$, that was derived from fits to the numerical simulations of \cite{grnzliss90, grnzliss92} instead of directly from N-body integrations.  P96 took the planetesimal velocities to be of the order of 1\,\kms and computed the 2-body trajectory of the planetesimal through the envelope for a series of impact parameters with the protoplanet. Once inside the envelope, the effect of gas drag on the planetesimal's trajectory and ablation was computed using a code developed by one of us (MP) \citep{podolak88}.   The papers cited in the first paragraph have also used this code for computing the interaction of planetesimals with the ambient gas as well as the resulting gas drag and planetesimal ablation.  Similar codes have been used in related studies, \cite[see, e.g.][]{alibert05}. 

If, after the encounter with the envelope, a planetesimal lost enough energy, or if it broke up due to ram pressure, P96 considered the planetesimal to have been captured and accreted onto the planet. By computing the trajectories for a series of impact parameters, the effective capture cross section of the protoplanet was calculated as a function of time and that, together with the planetesimal flux, gave the capture rate, and the resulting mass accretion rate.

\cite{inaba03b} revisited this part of the calculation using analytic fits to the N-body simulations of \cite{ida1993}, \cite{nakazawa1989}, and \cite{inaba2001}. They also used an analytic approximation for the structure of the protoplanetary envelope.  They computed two-body trajectories for the planetesimals in the field of the protoplanet, and ignored the effect of the central star inside the Hill sphere.  The ablation of the planetesimal once it entered the envelope was computed using an ``ablation factor" which does not depend on temperature. Finally, they argued that ram pressures were generally not high enough to cause planetesimals to break up.  They found that this procedure gives a higher accretion rate and a faster formation time than that found by P96. 

A related issue involves the inference, based on Juno gravity data, that Jupiter's heavy element core does not have a sharp outer boundary.  Rather, the heavy element mass fraction varies with radius \citep{wahl2017, debras2019}.  This raises the question of where in the proto-envelope the accreted planetesimals deposit their mass.  \cite{iaros07} originally explored this question using the data from the P96 model.  \cite{lozovsky17} revisited this question using an updated version of the P96 code. As a reminder, in the P96 model most of Jupiter's core is accreted before the envelope undergoes collapse, and it is difficult to mix the heavy material back into the envelope.  Recently, \cite{liu2019} have suggested that the original core material was later mixed by a giant impact, and that $\sim 10\,M_{\oplus}$ of heavy elements in Jupiter came from the impacting body.  Once again, the details of the planetesimal accretion play a central role in unraveling the details of Jupiter's formation, structure, and composition.

In their original calculation, P96 assumed that the planetesimals had a radius of 100\,km and were composed of a mixture of rock, water ice, and some generic organic material made up of a combination of carbon, nitrogen, oxygen, and hydrogen.  Although this is expected to be a typical composition for planetesimals, one would expect that planetesimals with different compositions and sizes would show a different response to gas drag and ablation, and would have correspondingly different trajectories through the gaseous envelope.  As a result, the accretion cross section and mass accretion rate might be expected to depend on planetesimal size and composition.

P96 also ignored the effect of the Sun's gravitational force in computing the trajectories of the planetesimals through the envelope.  Although the encounters with the protoplanetary envelope occur well within the Hill sphere of the protoplanet, and the Sun's gravity should have only a small effect on the planetesimal's trajectory through the envelope, the Sun will affect those planetesimals that are not captured and re-encounter the planet at a later time.  

In order to include these effects and investigate the importance of doing a more detailed computation for the motion of the planetesimals and their interactions with the gas during the contraction of the envelope, we have extended the code of \cite{podolak88} to explicitly compute planetesimal trajectories that are subject to the gravitational forces of both the Sun and Jupiter as well as to the gas drag forces that the planetesimal encounters in its motion through the envelope of the evolving planet.  As the focus of this study is on developing a more accurate picture of the effect of envelope gas drag, we focus our study on the three-body problem above and do not include additional objects orbiting the Sun. The effects of additional bodies, including Saturn, will be presented in future studies.  In what follows we describe this Explicit Solar System Trajectory Integrator (ESSTI) in more detail, and present the results obtained for the case of a growing Jupiter at 5.2 AU from the Sun. 

In addition to the gravitational interactions between the Sun, the protoplanet, and the planetesimal, there is also the interaction of the planetesimal with the nebular gas prior to its encounter with the protoplanet envelope.  This depends on the details of the  relative motions of the gas and the planetesimal.  Our primary objective in this work is to test the assumptions of P96 regarding the flux of planetsimals and their encounter speeds with the protoplanet, and to see how the differences can affect the accretion rate of the planetesimals.  Since the calculations of P96 ignored the effect of nebular gas on the planetesimals, we do so too for the initial calculations.

In Section 2 we discuss the numerical integrator, and in Section 3 we examine the effects of a more detailed computation of the planetesimal trajectories, both outside and inside the protoplanetary envelope.  In Section 4 we discuss the effects of nebular gas drag, and present some calculations showing how our results might be altered when a detailed model of nebular gas motion is included. In Section 5 we give our conclusions.  

\section{Planetesimal Properties and Trajectories}
In this study we consider the restricted, three-body system of the Sun, Jupiter and a planetesimal of negligible mass. We solve the equations of motion of the system in the barycentric coordinates where Jupiter and the Sun revolve around the center of mass of the system in nearly circular orbits. We consider the planetesimal to be subject to the gravitational forces of the Sun and Jupiter. However, we ignore its effect on the latter bodies due to its small mass.  Jupiter is treated as an extended mass with the radius and mass varying with time according to model A of \cite{lozovsky17} as described in the next section.

The equations of motion are integrated using a fourth-order Runge-Kutta integrator with adaptive step size. We implemented this integrator using both the Fehlberg45 \citep{fehlberg1969} and the Cash-Karp \citep{cash1990} methods.  Both methods are embedded algorithms of fourth-order that compute a fifth-order error estimate which allows for an automatic step-size correction.  As might be expected, the results were comparable, but in practice the Felberg45 method was slightly more efficient and was used for most of the computations.

The equations of motion are integrated in two stages. As long as a planetesimal is outside the protoplanet, the three body system of Sun, protoplanet, and planetesimal is integrated, as explained above, and Jupiter is treated as a point mass.  When the distance between a planetesimal and Jupiter's center of mass becomes less than Jupiter's radius at the time, we consider the planetesimal as having entered the gaseous envelope of the protoplanet.  In this case the gravitational force exerted on the planetesimal by the Sun is computed as before, but the gravitational force of the protoplanet is due to the mass between the planetesimal and the protplanet's center of mass.  In addition, the gas drag force is computed using the prescription given by \cite{podolak88} and is added to its equation of motion.  Further details are given in the appendix.  As noted above, the gas drag not only affects the trajectory of the planetesimal but also heats it.  In addition, the planetesimal is also heated by radiation from the ambient gas.  We compute the mass loss due to ablation caused by these two mechanisms.  Another process that contributes to the accretion of a planetesimal is the ram pressure. If the ram pressure that is experienced by a planetesimal is greater than its tensile (material) strength, the planetesimal will fragment \citep{poletal79}. Because the ram pressure is caused by gas drag, the efficiency of this fragmentation varies with the size of a planetesimal and its velocity relative to the gas.  If the ram pressure is sufficient to break up the planetesimal, we take the planetesimal to be accreted.

We considered three planetesimal sizes: 1, 10, and 100\,km in radius.  For each of the three sizes we considered, in turn, three different compositions: ice ($\rho=1.0$\,\gcc), rock ($\rho=3.4$\,\gcc), and a mixture of 30\% ice and 70\% rock by mass ($\rho=2.0$\,\gcc).  This latter density is similar to that of objects like Titan, Triton, and  Pluto.  Further details of the physical parameters assumed are given in Table 1 (below) and in Table 2 of the appendix (see also Tables 1 and 2 of \cite{podolak88}).

\begin{table}[h] \centering
	\begin{tabular}{|c|c|c|c|}\hline
		Parameter &  Ice    &   Rock  & Mixed \\ \hline
		Density (\gcmc)   & 1.0 & 3.4 & 2.0 \\
		$P_0$ (dyn\,cm$^{-2}$)   & $3.891\times 10^{11}$ & $1.50\times 10^{13}$  & $3.891\times 10^{11}$ \\
		$A$ (K) & $2.1042\times 10^3$ & $2.4605\times 10^4$  & $2.1042\times 10^3$ \\
		Tensile Strength (dyn\,cm$^{-2}$)  & $10^6$  & $10^8$ & $10^6$ \\ \hline
		\hline
	\end{tabular}
	\caption{Physical parameters assumed for the planetesimals.  $P_0$ and $A$ are the parameters assumed in computing the vapour pressure according to the formula $P(T)=P_0 e^{-A/T}$ where T is the temperature at the planetesimal surface.}
\end{table}  

We chose these three compositions in order to explore different regimes of the planetesimal-gas interaction.  The ice planetesimals are volatile and have low density.  Not only do they heat up and evaporate easily, their trajectories are strongly influenced by the gas drag.  The rock planetesimals have a much higher density, and are less strongly affected by gas drag.  As a result, they suffer significantly less ablation. In addition, their tensile strength is two orders of magnitude higher so they are much more stable against ram-pressure break-up.    

The mixed ice-rock planetesimals are the more realistic, intermediate case.  Here we view the planetesimal as composed of small rock particles embedded in an ice matrix.  As the ice evaporates, it carries embedded rocky particles with it as well.  This means that the mixture has essentially the same volatility as ice. However, since the planetesimal has an overall density intermediate between rock and ice, we can explore the effect of higher density coupled with high volatility. 

We considered approximately 2000 planetesimals, with small ($\leq 0.05$) random initial eccentricities and with semimajor axes randomly distributed between 3.7 AU and 6.7 AU, corresponding to four Hill sphere radii on either side of the planet. In one set of integrations (``set I"),  we considered no inclination for the orbits of all planetesimals and the orbit of Jupiter, and in a second set (``set II"), we considered orbital inclinations randomly varying between 0 and 0.003 deg.  The integration was stopped and the planetesimal was considered accreted if one of three outcomes occurred: the planetesimal hit the core\footnote[1]{A collision is taken to occur when the distance between the centers of mass of the two bodies becomes equal to the sum of their radii. }, the mass loss due to ablation was greater than 80\% of the original planetesimal mass, or the planetesimal broke up due to ram pressure.  If the planetesimal left the envelope, any ablated material was assumed to be left behind and was added to the total mass accreted by the protoplanet.  In this case the planetesimal continued with its new mass and radius until it re-encountered the planet or the simulation ended.  We ran the simulations for $3\times 10^6$\,yr, which is roughly twice the time required for the protoplanetary envelope to collapse (see below).

\section{Effect of Gaseous Envelope}

\subsection{Models of the Protoplanetary Envelope Evolution}
\begin{figure}
\centerline{\includegraphics[width=20cm]{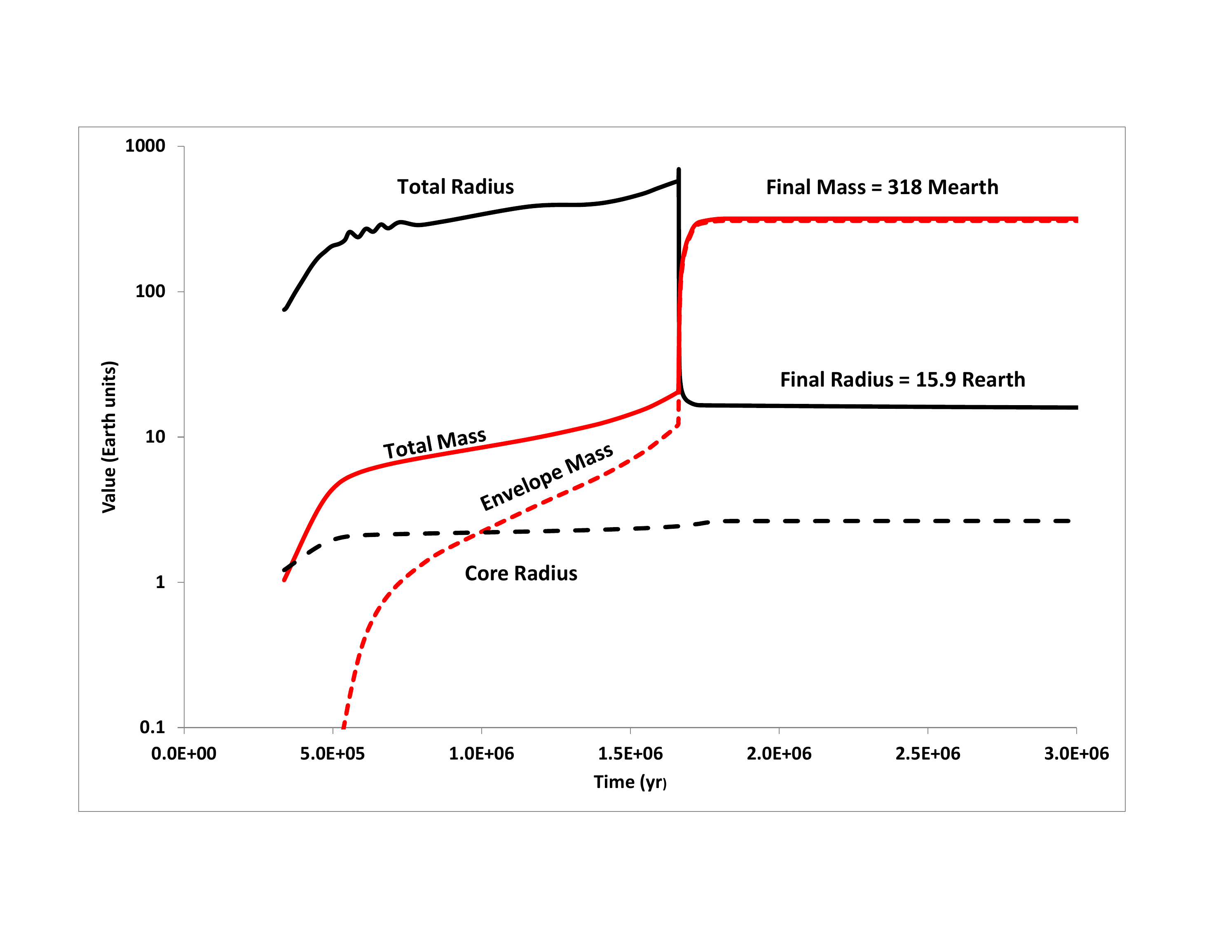}}
\caption{Planetary parameters used for the CA case.  The black curves show the core radius (dotted) and total radius of the protoplanet (solid) in Earth radii as a function of time.  The red curves show the total mass (solid) and envelope mass (dotted) in Earth masses as a function of time.}
\label{pmodel}
\end{figure}

In the CA scenario, the core of the protoplanet forms by accreting solid material from its surroundings.  This core, in turn, attracts gas from the surrounding nebula to form the protoplanet's gaseous envelope.  The combined mass of the core and gas increases slowly, and when the masses of the core and the gaseous envelope are roughly equal, the internal pressure of the gas is no longer able to withstand its gravity, and the gas collapses to a more compact hydrostatic equilibrium state.  The structure of the planetary envelope is determined by computing a series of models using the standard code as described in P96 and updated by \cite{lissetal09}.  Alternative calculations of planetesimal accretion rates from those of P96 are presented by \cite{zhou2007} and \cite{shiraishi2008}. They result in different scenarios for giant planet formation from that of P96. We will explain these studies in detail at the begining of Section 4.  As a representative case we take the time variation of the radius of the protoplanet and its internal density distribution from Model A of the latest series of such models as described in \cite{lozovsky17}.  This model follows the growth of a planet in the CA scenario for an assumed background surface density of solids of $\sigma=6$\,\gcms\, and a planetesimal radius of 100\,km.  Figure \ref{pmodel} shows the time variation of the mass and radius for this evolution. Shown are the total radius (solid black), the core radius (dotted black), the total mass (solid red) and the envelope mass (dotted red).  The protoplanet begins with approximately a Mars mass at time zero.  Thereafter the mass increases linearly with time until a core of approximately 1 Earth mass forms at $3.36\times 10^5$\,yr.  After that the core grows as shown in the figure.  There is a rapid growth during the first half-million years, after which the growth slows while a gaseous envelope is accreted.  At around $1.7\times 10^6$\,yr the envelope collapses and the mass increases to the current Jupiter mass in around $10^5$\,yr.  During the evolution, the protoplanet's radius first increases from $10^2$ to $10^3$ Earth radii and drops quickly during the collapse stage to around 15 Earth radii. It then slowly cools to the present day Jupiter radius.  The above described evolution was based on an assumed mass accretion rate that was determined in the manner of P96.

As noted in the introduction, \cite{inaba03b} also investigated the problem of planetesimal accretion.  However, based on their calculations, \cite{inaba03b} argue that the accretion rate used by P96 is too low.  In order to investigate this claim, we performed the following numerical experiment:  Model A, which was computed using the accretion rate of P96 was used to describe the structure of the protoplanet as a function of time.  We then used the ESSTI code to compute detailed trajectories of the planetesimals, including the gravitational forces of both the Sun and Jupiter, and the gas-drag forces, and ablation caused by the gaseous envelope.  In this way, we computed a more realistic accretion rate than that estimated by either P96 or \cite{inaba03b}.  If this accretion rate would be similar to the P96 rate originally assumed in calculating model A, then that model would be self-consistent.  Otherwise the P96 estimate and the resulting Jupiter evolution as given, for example, by the models in \cite{movsh10} and \cite{lozovsky17} would need to be revised.

\begin{figure}
\centerline{\includegraphics[width=14cm]{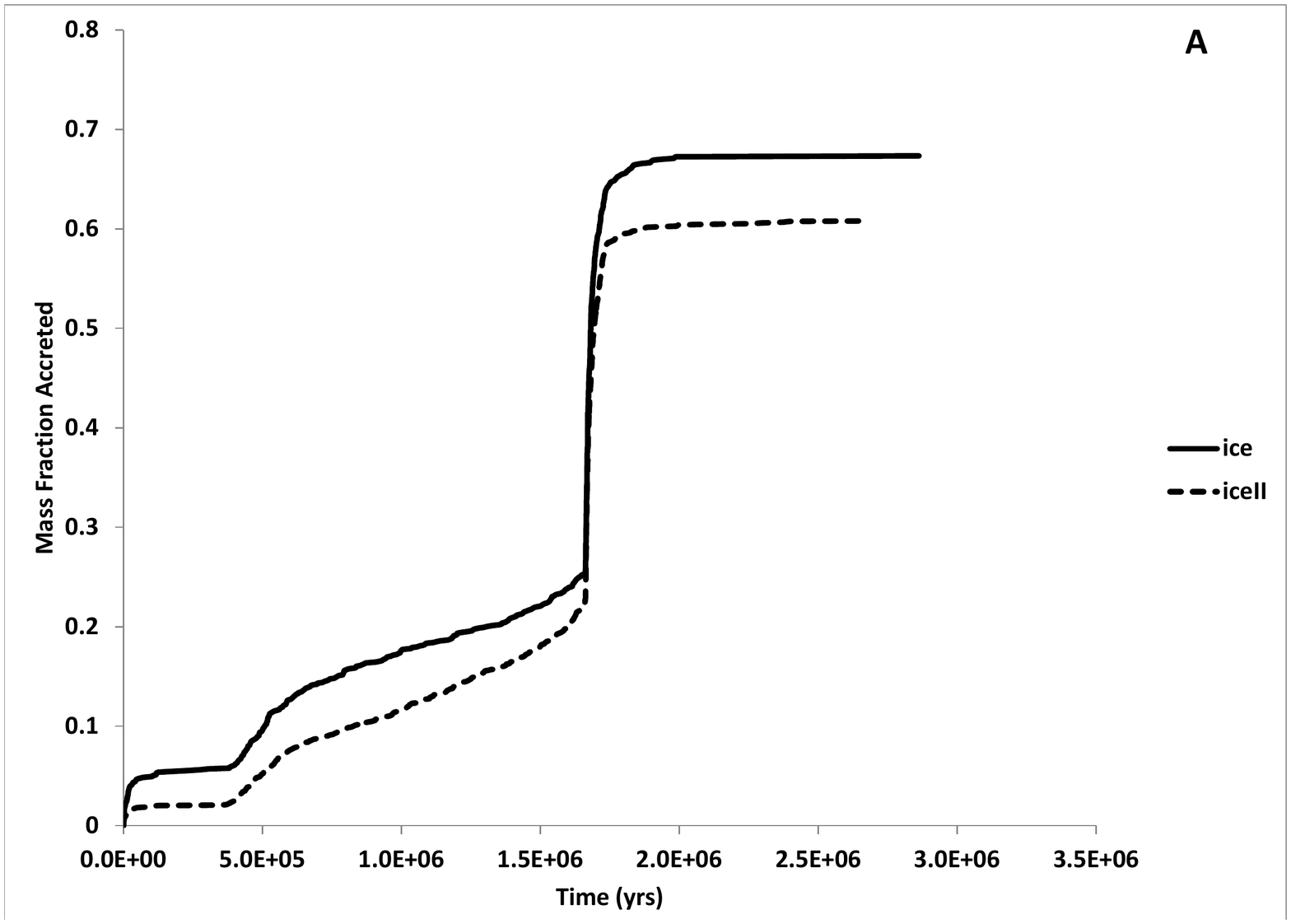}}
\centerline{\includegraphics[width=14cm]{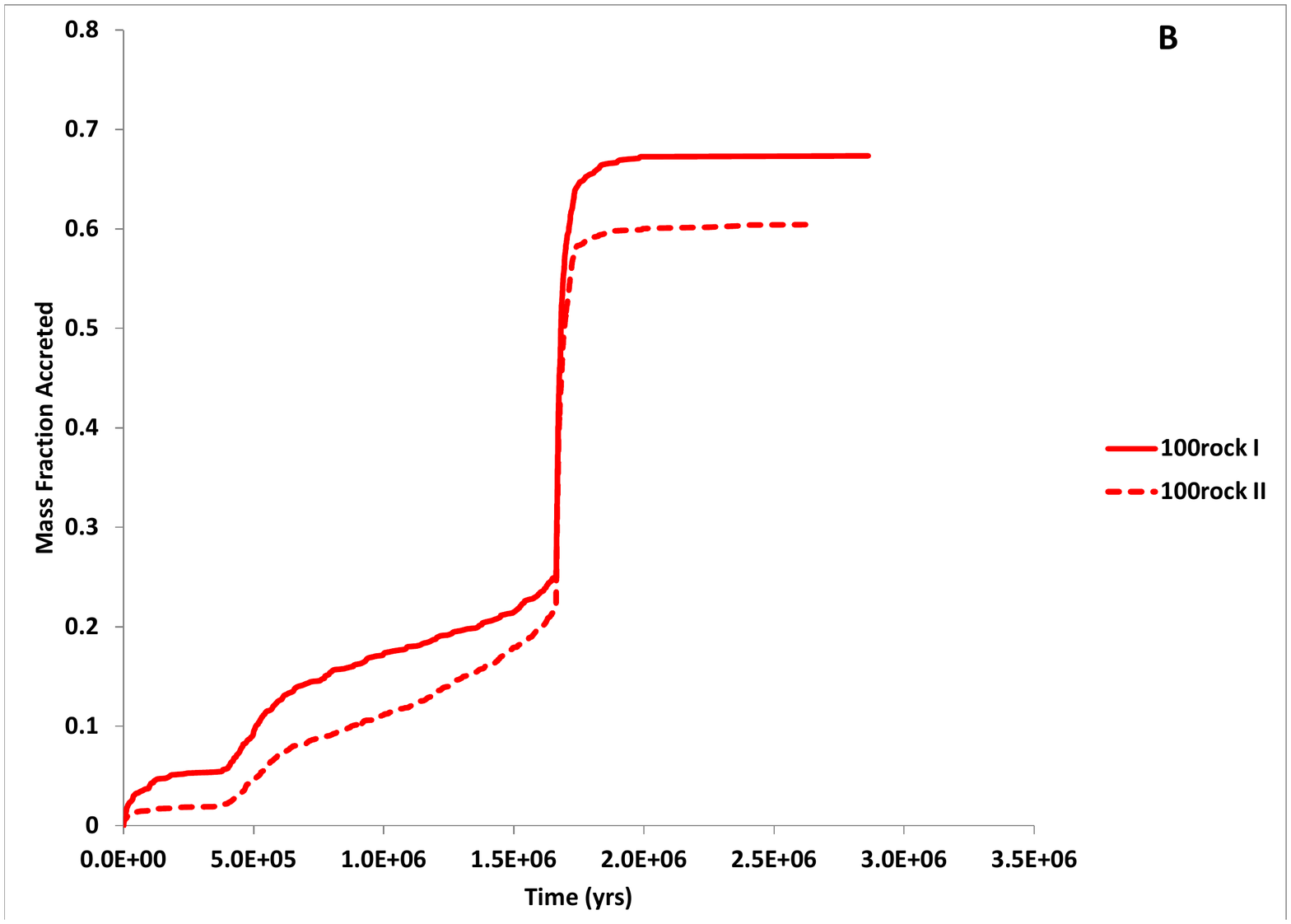}}
\caption{Mass fraction of planetesimals accreted as a function of time.  Upper panel (black curves): 1\,km ice planetesimals of set I (solid), and set II (dotted).  Lower panel (red curves): 100\,km rock planetesimals of set I (solid) and set II (dotted).}
\label{maccp}
\end{figure}

Figure \ref{maccp} shows the cumulative fraction of the available planetesimal mass between 3.7 and 6.7\,AU that is captured and accreted onto the protoplanet as a function of time.  Four curves are shown.  The solid curves are for planetesimals of set I, and the dashed curves are for set II planetesimals.  The black curves are for 1\,km ice planetesimals and the red curves are for 100\,km rock planetsimals.  The accretion rate for set II is initially lower than that of set I because Jupiter's relatively small cross section at this time means that if a planetesimal is even slightly out of the orbit plane of the protoplanet, the chance for it to be accreted becomes small.  

Figure \ref{maccp} also shows that for a given set of orbital parameters, the accretion probability is nearly independent of planetesimal size and composition.  While this result may seem counter-intuitive, it can be understood as follows.  The planetesimals that enter the protoplanetary envelope have a typical encounter speed of $\sim 5$\,\kms  largely as a result of falling into the protoplanet's gravitational well.  At this speed, if the gas density is $\gtrsim 10^{-5}$\gcc, ice planetesimals will fragment due to the ram pressure of the gas.  In comparison, the 100\,km rock planetesimals have a compressional strength that is roughly two orders of magnitude larger, however they also have more inertia and enter the region where the gas density approaches $10^{-3}$\gcc  with a velocity that is a factor of two or more higher than the 1\,km ice planetesimals.  As a result, the ram pressure, which scales with the gas density and the square of the velocity is correspondingly higher, and breakup still occurs.  Thus a 1\,km ice planetesimal and a 100\,km rock planetesimal in identical orbits will generally (but not always!) have similar fates.  They will either not encounter the protoplanet at all, in which case they will both continue on identical orbits, or they will both encounter the protoplanet.  In the latter case, although the gas drag effects will be very different for the two bodies, they will, in most cases, both be accreted.

\subsection{Effect of the Protoplanetary Envelope - An Example }
It is important to note that, although the envelope initially extends to tens or hundreds of Earth radii, the gas density in the outer regions is very low, especially in the first few $10^5$ years of growth.  As a result, during this early period, although the radius of the envelope is very large, it is only in the innermost regions where gas drag has sufficient effect, especially on the smaller bodies, to contribute to planetesimal capture.  This leads to an effective capture cross section that is only a few times larger than the geometric cross section of the core. Indeed most captures result from planetesimals simply colliding with surface of the protoplanetary core, or in the case of small icy planetesimals, breaking up in the envelope just above the core.  Since the 1\,km ice planetesimals lose more energy due to gas drag than their denser and less volatile rocky counterparts, they are captured somewhat more quickly in the very beginning. This can be seen in Figure \ref{maccp} where the black solid curve initially rises more quickly than the red solid curve.  The effect exists for the dashed curves (set II) as well, but is less pronounced.  As time goes on and the protoplanetary envelope becomes denser, gas drag begins to affect the larger bodies as well, and eventually most of the bodies that began the simulation in the immediate vicinity of the protoplanetary core are accreted by it.  

\begin{figure}
\centerline{\includegraphics[width=14cm]{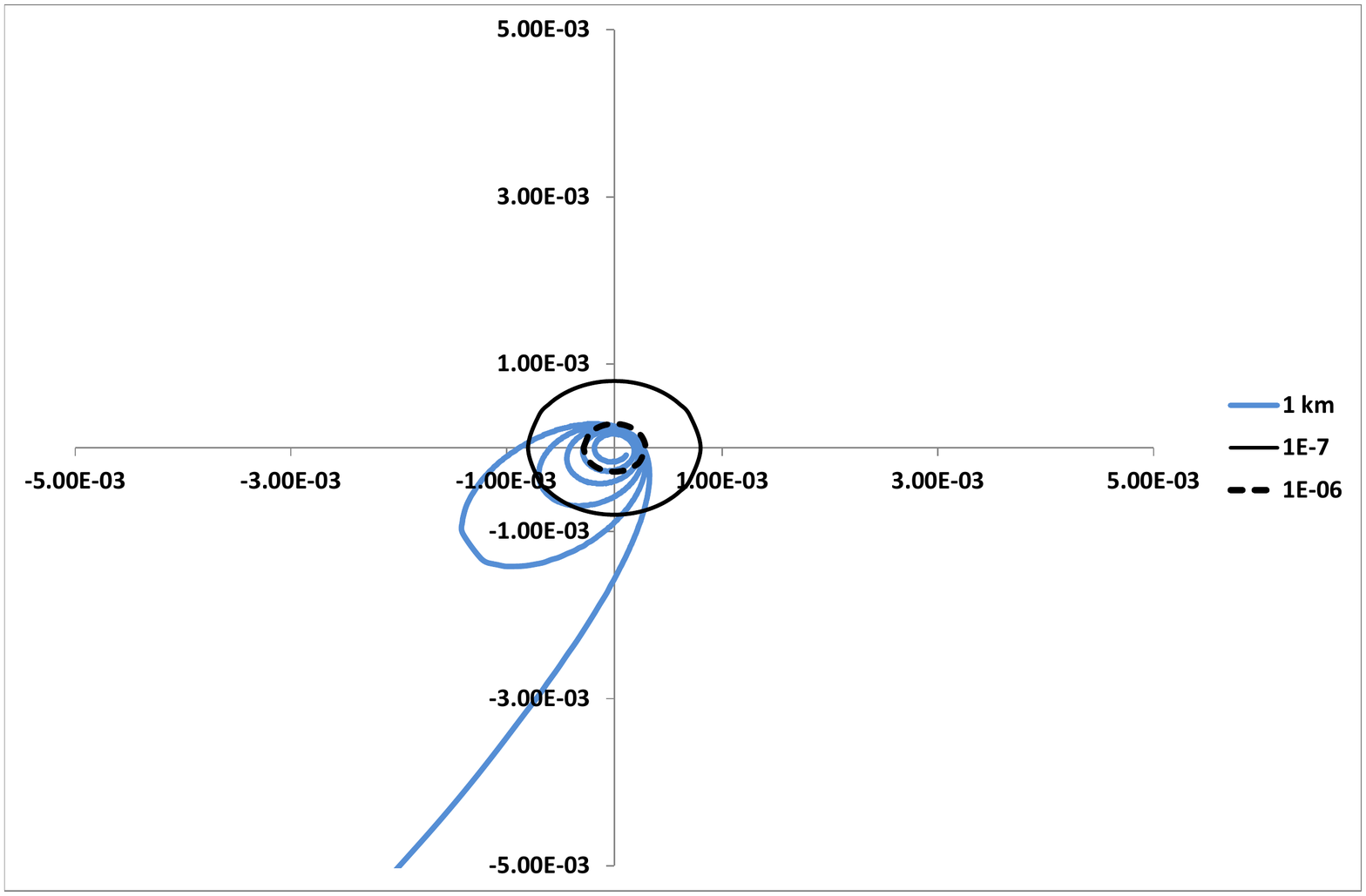}}
\centerline{\includegraphics[width=14cm]{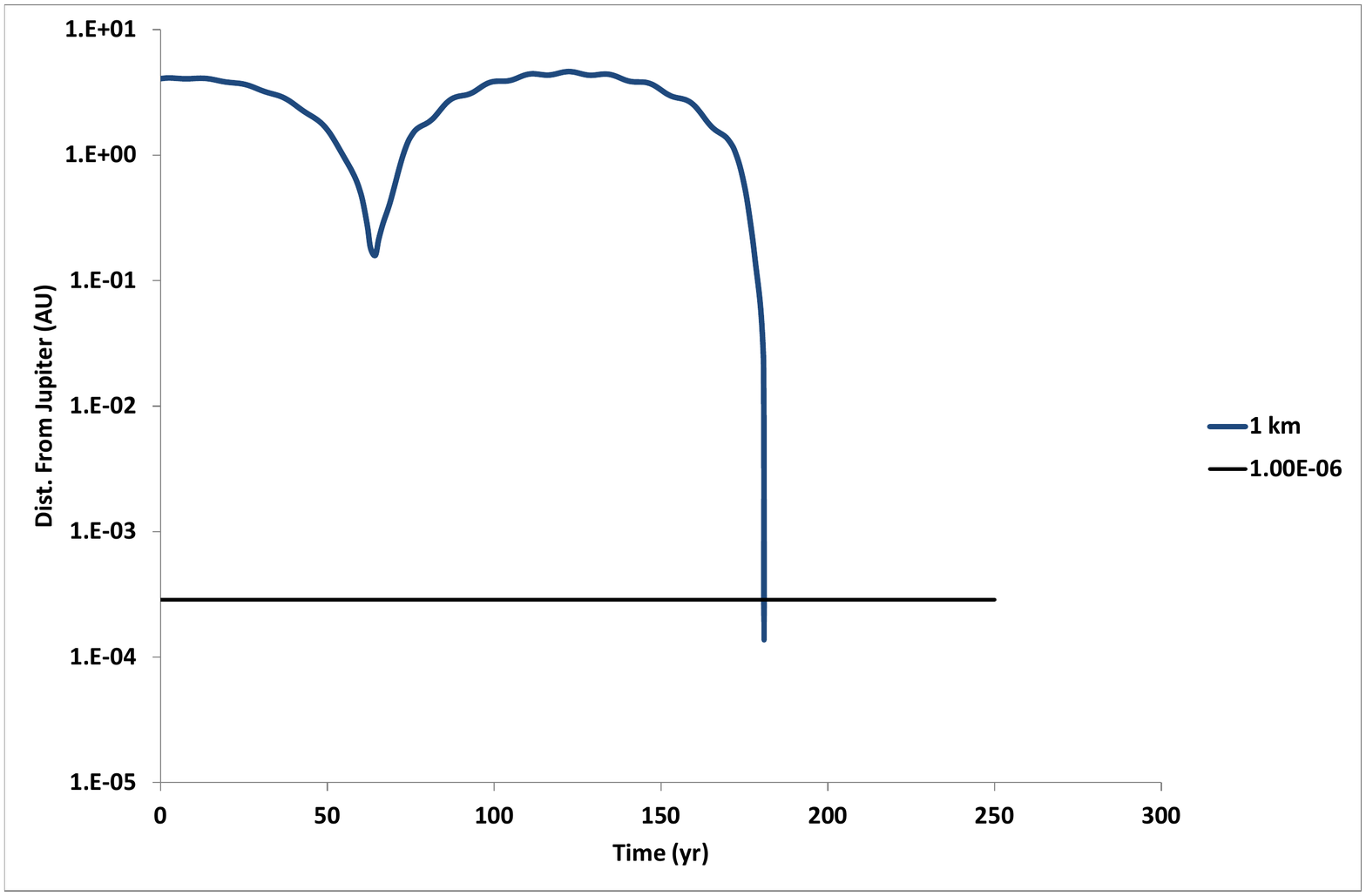}}
\caption{Upper panel: Trajectory of 1\,km ice planetesimal around the protoplanet (blue curve).  Outer black circle marks the region where the density of the envelope is $10^{-7}$\,\gcc\ and the inner dashed circle marks the region where the density is $10^{-6}$\,\gcc. Lower panel: Distance (in AU) of planetesimal from protoplanetary center as a function of time.  The horizontal black line marks the level in the envelope where the envleope gas density is $10^{-6}$\,\gcc.}
\label{orbit1a}
\end{figure}

\begin{figure}
\centerline{\includegraphics[width=14cm]{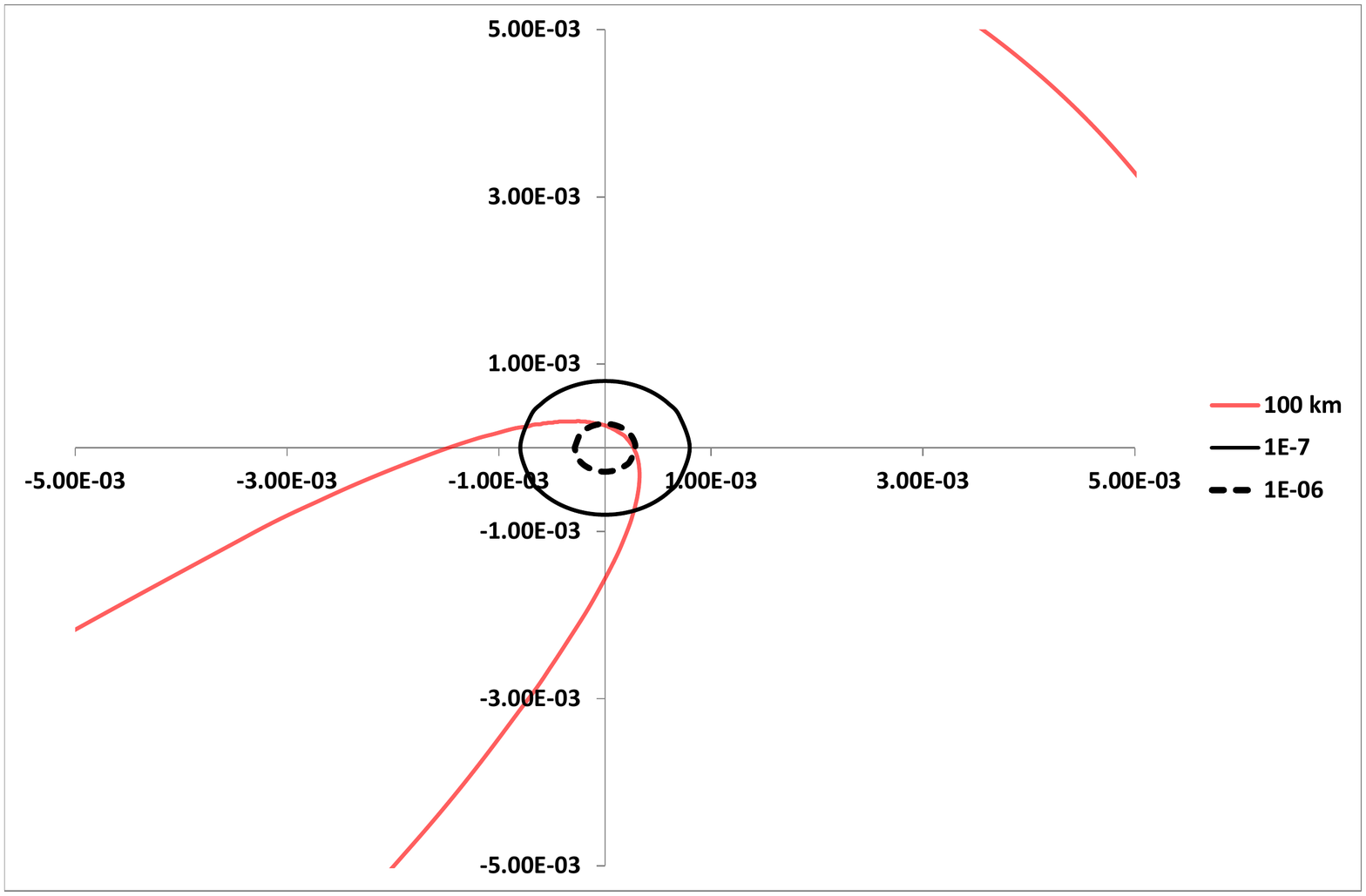}}
\centerline{\includegraphics[width=14cm]{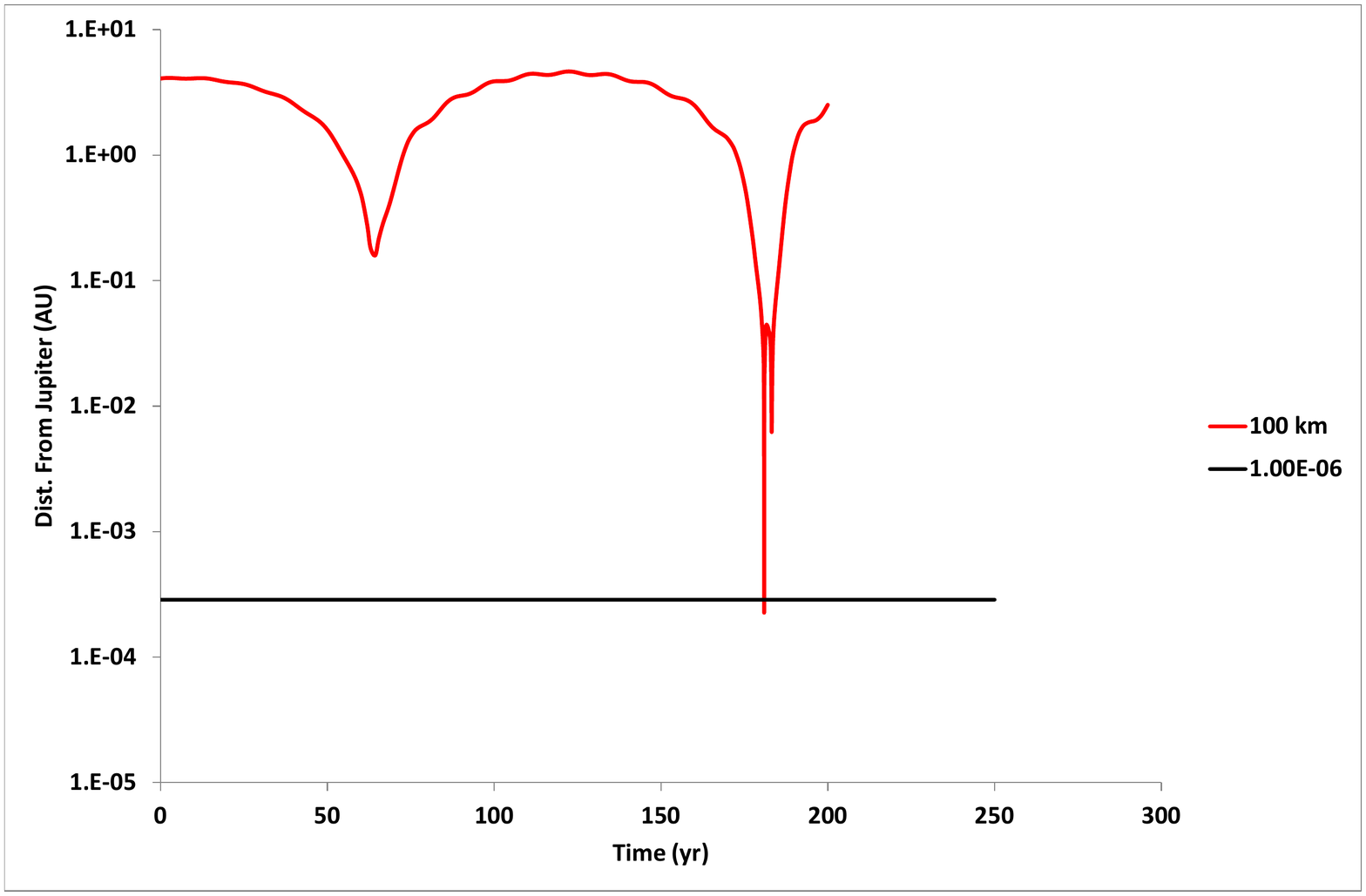}}
\caption{Same as Figure\,\ref{orbit1a} but for a 100\,km rock planetesimal (red curve). Distances are in AU.}
\label{orbit1c}
\end{figure}

An example of this behavior, is a planetesimal of set I with an initial semi-major axis of 5.22\,AU and an eccentricity of $7.39\times 10^{-3}$.  Figure \ref{orbit1a} shows the first 200 years of motion for a 1\,km ice planetesimal.  The upper panel shows the actual trajectory of the planetesimal with respect to the protoplanet which is at the center. Here the solid black circle around the center marks the region in the protoplanetary envelope where the gas density reaches a value of $10^{-7}$\,\gcmc, and the dashed circle marks the region where the gas density reaches a value of $10^{-6}$\,\gcmc.  As a point of comparison, for typical encounter velocities, ice planetesimals generally break up as a result of ram pressure when they enter a region where the gas density is a few times $10^{-5}$\,\gcmc, while rock planetesimals break up when they enter a region where the gas density is closer to $10^{-3}$\,\gcmc. The lower panel shows the distance of the planetesimal from the center of the protoplanet as a function of time.  Figure \ref{orbit1c} shows the case for a 100\,km rock planetesimal with the same initial orbital parameters.  

\begin{figure}
\centerline{\includegraphics[width=12cm]{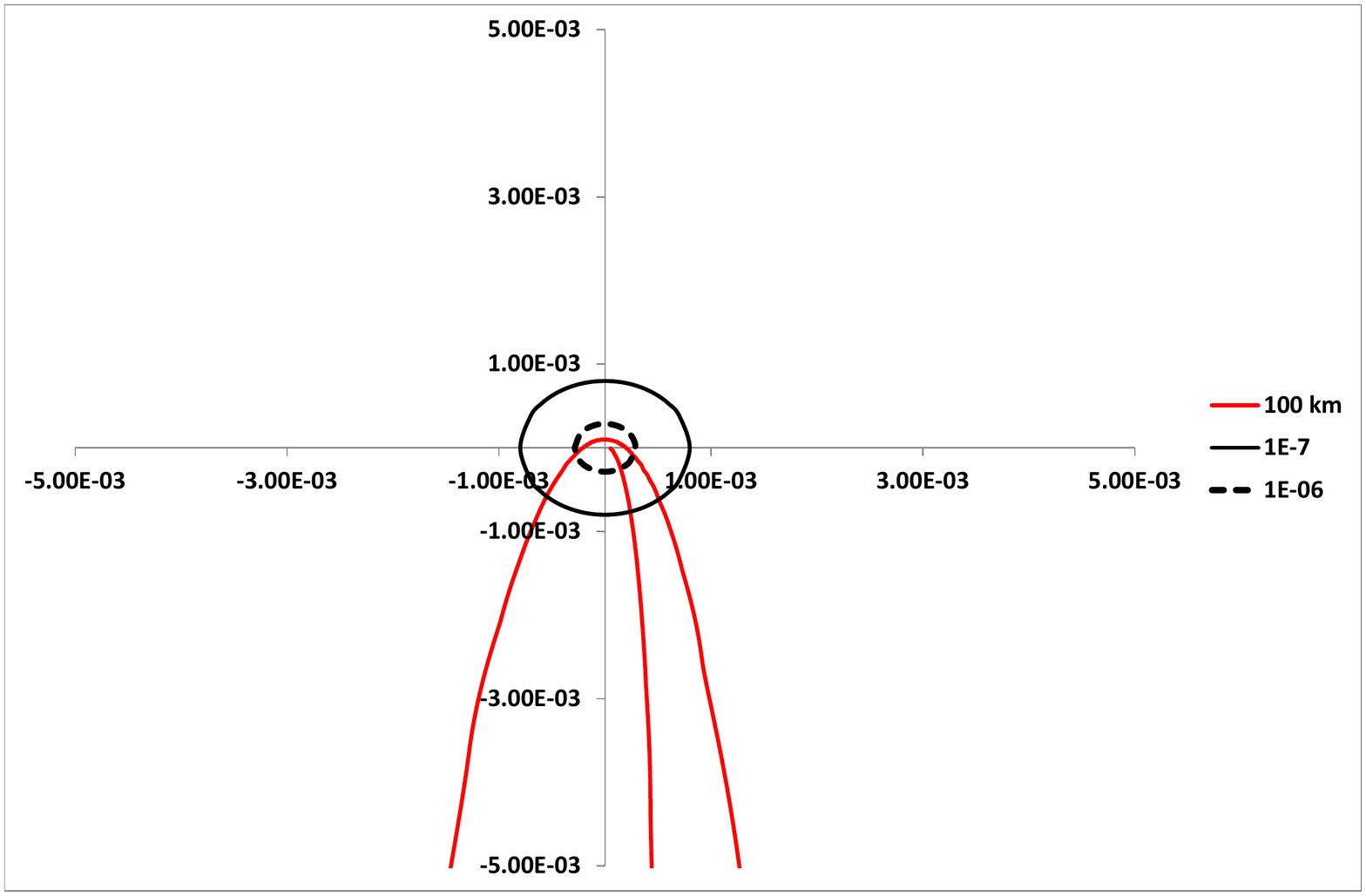}}
\centerline{\includegraphics[width=10cm]{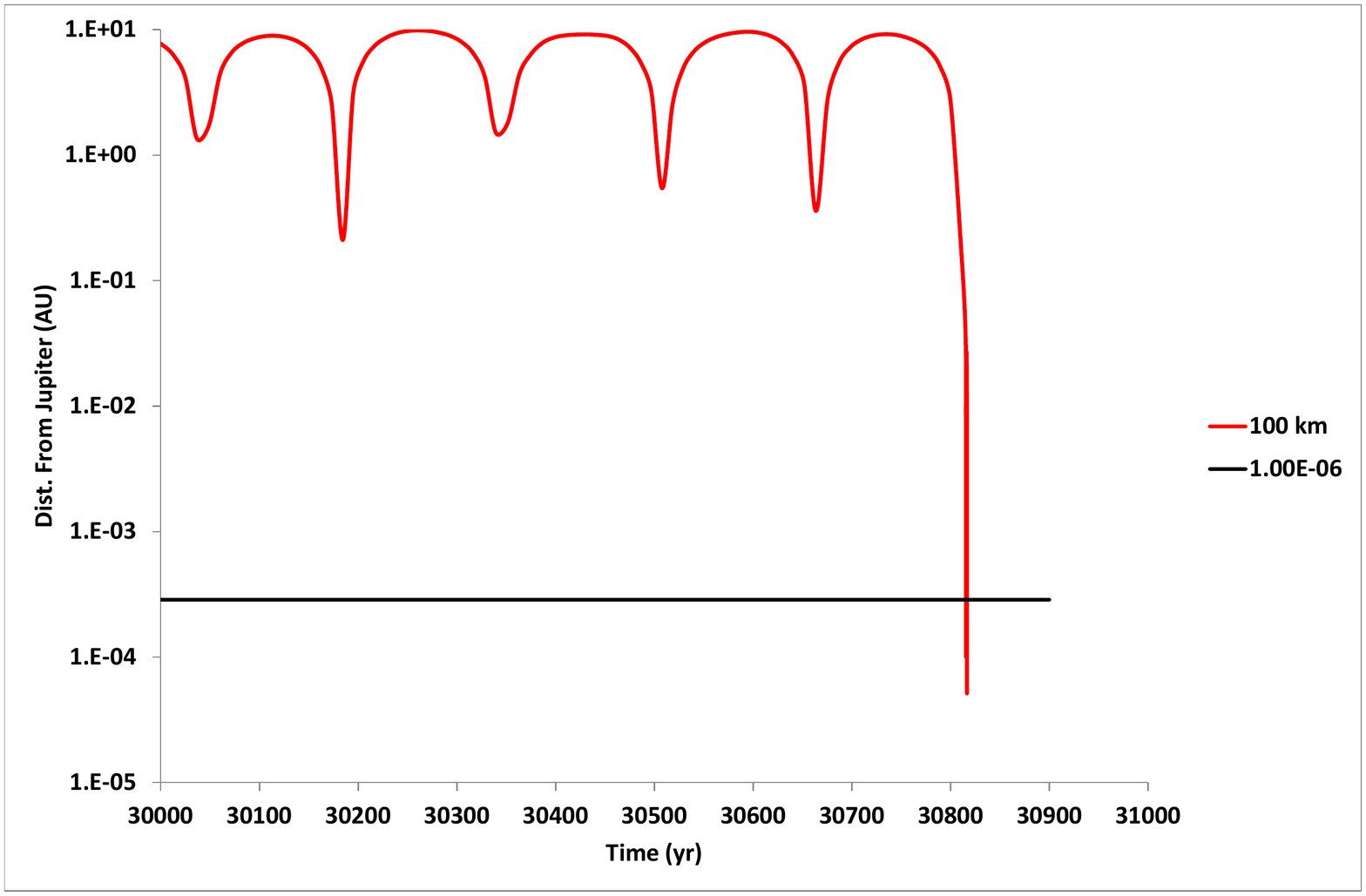}\includegraphics[width=10cm]{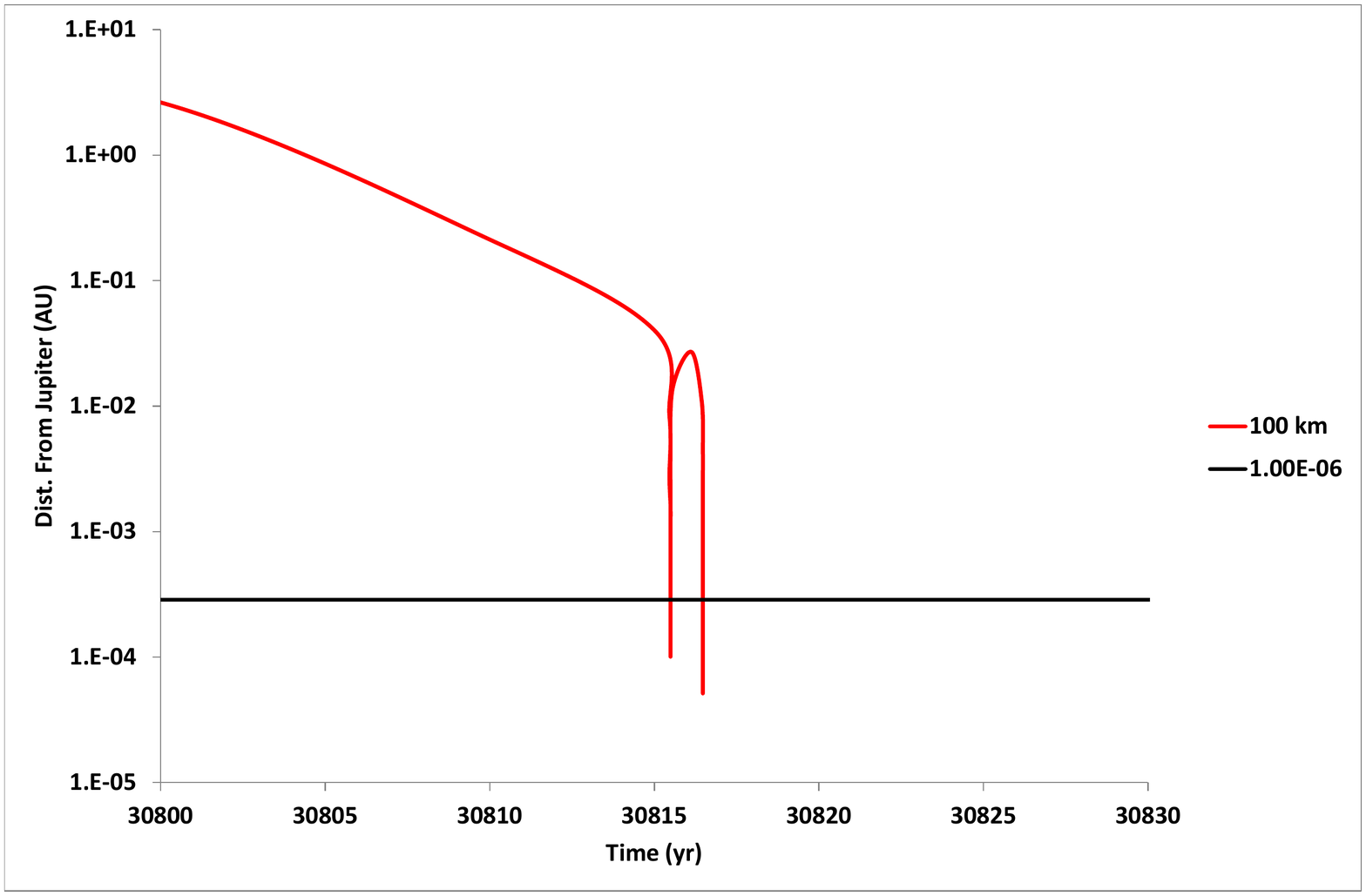}}
\caption{Encounter of 100\,km rock planetesimal with protoplanet after $\sim 31,000$\,yr.  Upper panel: Outer black circle marks the region where the density of the envelope is $10^{-7}$\,\gcc\ and the inner dashed circle marks the region where the density is $10^{-6}$\,\gcc.  Lower left panel: Distance (in AU) of planetesimal from protoplanetary center as a function of time.  Lower right panel: Same, but with an expanded time axis to show the period between 30,800 and 30,830\,yr.  In both lower panels the horizontal black line marks the level where the envelope gas density is $10^{-6}$\,\gcc.  }
\label{orbit2a}
\end{figure}

As can be seen from the figures, both bodies enter the gaseous protoplanetary envelope, which is just beginning to be accreted onto the core, after $\sim 180$ years.  The smaller body experiences sufficient gas drag during this encounter so that it no longer has the energy to escape the protoplanet and spirals into the deeper atmosphere, where it breaks up due to ram pressure some 13,000\,km above the surface of the core.  For the 100\,km rock case, the planetesimal's inertia is large enough so that it easily escapes and is only captured 31,000 years later when it re-encounters the protoplanet and enters deeply enough into the envelope so that it can be slowed by gas drag and captured.  This can be seen in Figure \ref{orbit2a}.  The upper panel shows the trajectory with respect to the protoplanet's center.  Here we see that the planetesimal has another close encounter with the protoplanet.  This time, however, it goes deep enough into the envelope so that it experiences gas densities of $\gtrsim 10^{-5}$\,\gcmc\ and the resultant drag is sufficient to prevent it from escaping the protoplanet's Hill sphere, and it eventually crashes onto the surface of the core.  The distance of the planetesimal from the protoplanetary core is shown in lower left panel, and on an expanded scale, for the 30 years around the time of capture, in the lower right panel.  The results of this and many similar cases indicate that although the details of the capture process may differ, the fraction of the available mass that is accreted as a function of time is very similar for different plantesimal sizes and compositions. 

\begin{figure}
\centerline{\includegraphics[width=20cm]{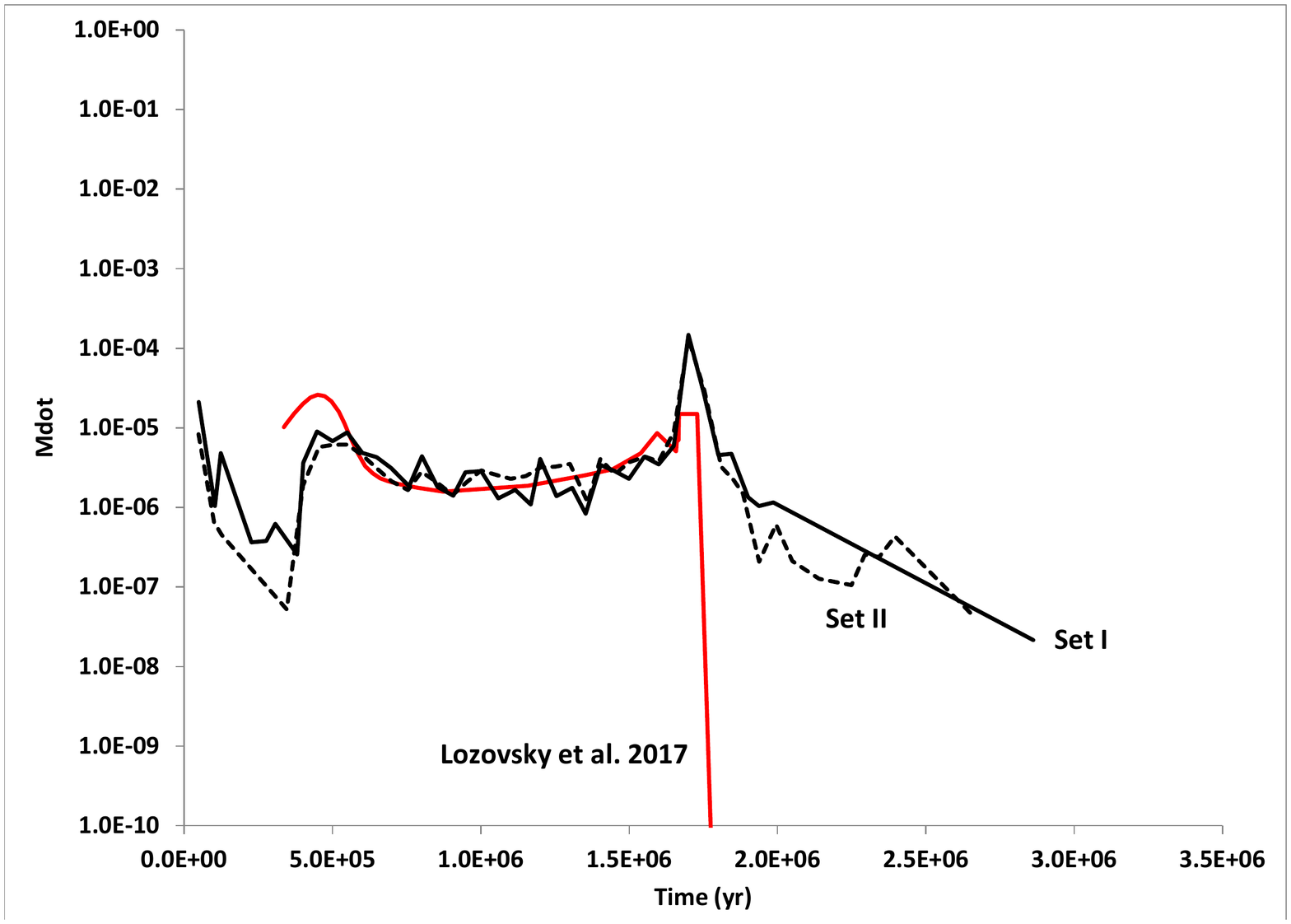}}
\caption{Mass accretion rate (in units of $M_{\oplus}$/year) onto the protoplanet as a function of time.  The red curve is the rate assumed by \cite{lozovsky17} in computing their models of the protoplanetary envelope.  The black curves are the accration rates derived from our model for set I (solid) and set II (dashed).  See text for details.}
\label{maccpcomp}
\end{figure}

If we assume, following \cite{lozovsky17} that the solid surface density at 5.2\,AU is 6\,\gcms, and that it has a radial profile of $r^{-3/2}$, then we can compute the total planetesimal mass expected between 3.7 and 6.7\,AU, and translate the fractional accretion rate of Figure \ref{maccp} into a mass accretion rate.  This is shown in Figure \ref{maccpcomp}, where, to reduce the fluctuations, we present the average mass accretion rate binned over intervals of $5\times 10^4$ years.  As can be seen, the accretion rates for cases I and II are qualitatively similar and show similar trend to that of \cite{lozovsky17} (red curve) up to the time of rapid collapse of the envelope.  However after the rapid increase in the mass of both the envelope and the protoplanet itself, the accretion rate is significantly higher than that assumed in the \cite{lozovsky17} model, and remains non-zero beyond the time of collapse, although at a much reduced rate. The red curve also has a small bump at around $3-5\times 10^5$ years.  

\begin{figure}
\centerline{\includegraphics[width=20cm]{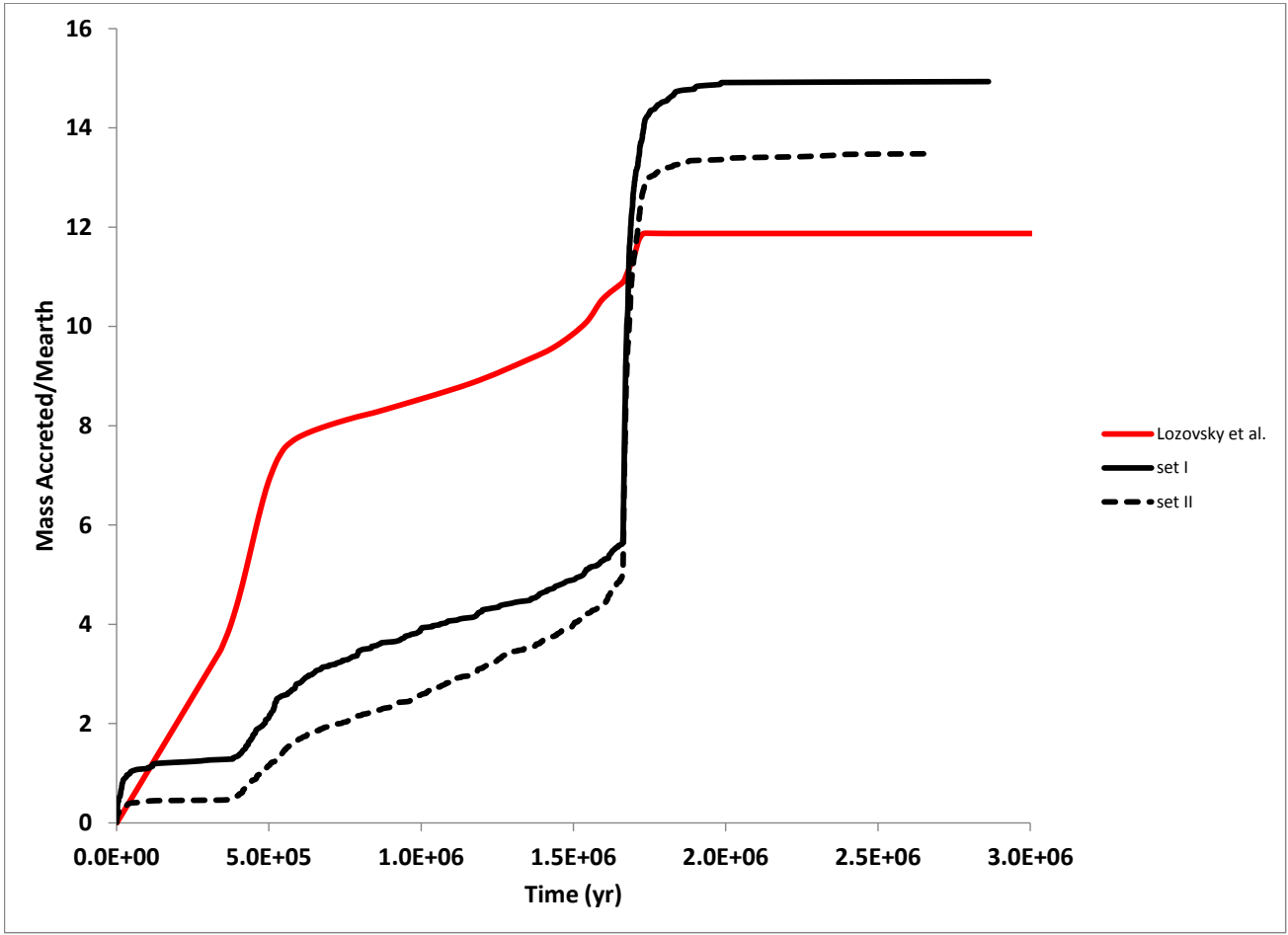}}
\caption{Accreted mass as a function of time.  The red curve is the mass accreted according to \cite{lozovsky17} and the black curves show the accreted mass as a function of time for planetesimals from set I (solid curve) and set II (dashed curve).  See text for details.}
\label{cormas}
\end{figure}

This has important consequences as can be seen in Figure \ref{cormas}.  The small bump in the mass accretion rate around $5\times 10^5$\,yr means that the accreted heavy element mass rises quickly in this region and, since this mass settles onto the core, the core mass increases quickly to around 8\,$M_{\oplus}$ at this time.  In our model, the accretion rate is lower and the core mass is only around 2\,$M_{\oplus}$.  As a result, most of the accreted heavy element mass goes into the core in the \cite{lozovsky17} model, whereas we find that most of the heavy element mass is accreted just around the time of the envelope collapse.  This, together with the fact that there is some additional accretion well after the collapse, should result in a much larger fraction of heavy elements remaining in the envelope.  Although the mass accretion found by \cite{d'angelo14} for the early stages of planet growth is notably higher than that computed here, there are two main reasons for this.  In the first place, \cite{d'angelo14} assumed $\sigma=10$\,\gcms\, for the surface density of solids, whereas we assume $\sigma=6$\,\gcms.  In addition, they included planetesimals as small as 10\,m in their model.  Bodies as small as this have their velocities relative to the protoplanet significantly damped by the nebular gas, and the resulting cross-section for their capture is correspondingly larger.  Finally, the cross-sections of \cite{d'angelo14} were computed using 2-body trajectories, and did not include the tidal effect of the Sun.

\begin{figure}
\centerline{\includegraphics[width=22cm]{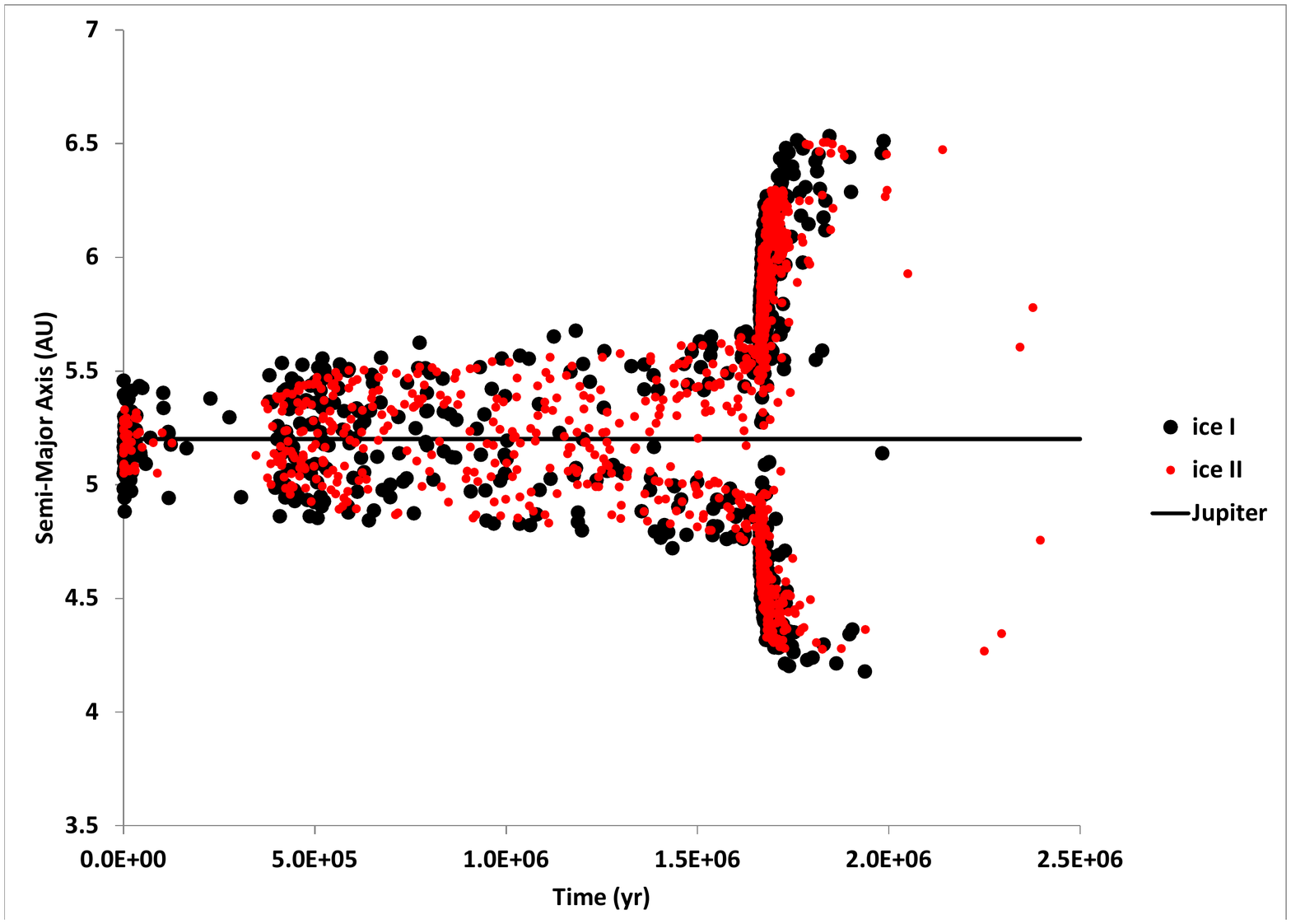}}
\caption{Initial semimajor axis of accreted planetesimals as a function of time for 1\,km ice planetesimals of set I (black dots) and set II (red dots). The position of Jupiter is shown by the solid black line.}
\label{tuning}
\end{figure}

Figure \ref{tuning} shows these effects from another point of view.  Here we plot the semimajor axis of an accreted planetesimal as a function of the time of its capture for 1\,km ice planetesimals of set I (black dots) and set II (red dots).  The horizontal straight line indicates the position of Jupiter.  As can be seen, at very early times planetesimals of set I are accreted from a larger radial region around Jupiter than planetesimals of set II.  As mentioned before, this is due to the small inclinations of the latter planetesimals which results in a smaller cross section for their accretion. As time progresses and the mass of Jupiter increases, the accretion cross section is enlarged and as a result, planetesimals at larger distances from both sets are more readily accreted.  

\begin{figure}
\centerline{\includegraphics[width=14cm]{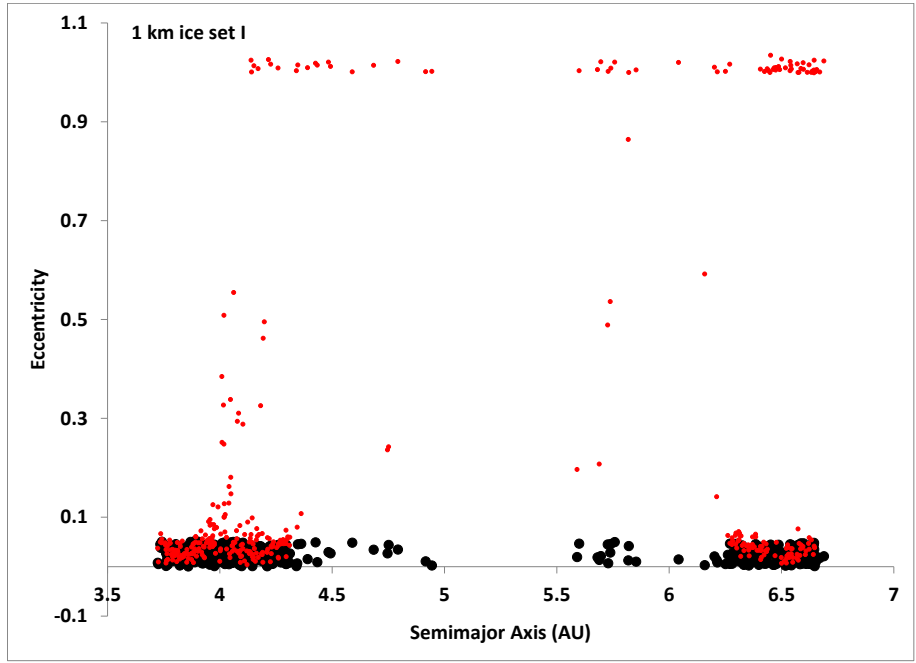}}
\centerline{\includegraphics[width=14cm]{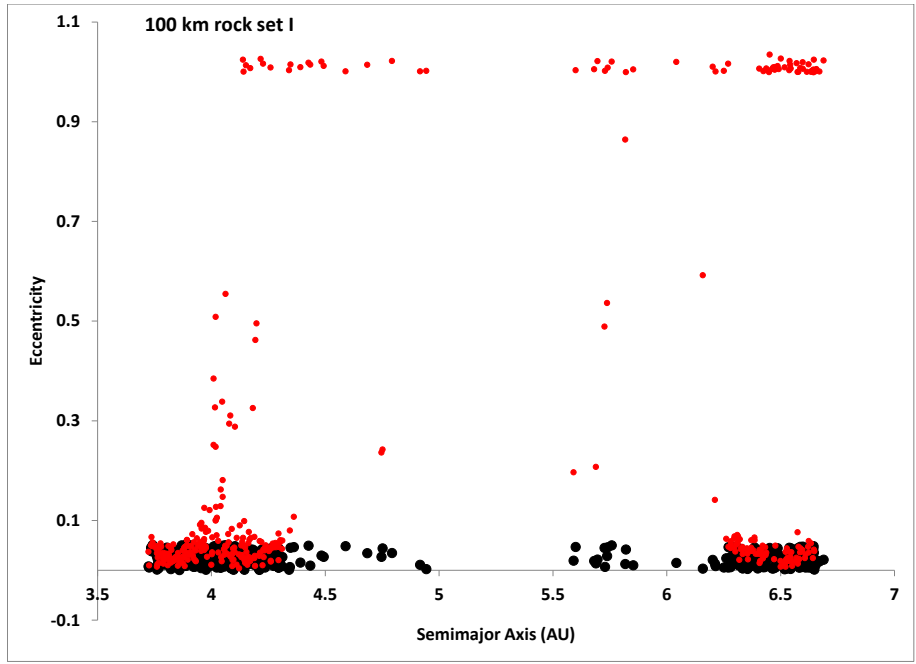}}
\caption{Upper panel: Initial eccentricity (black dots) and final eccentricity (red dots) of non-captured 1\,km ice planetesimals of set I as a function of initial semimajor axis. Lower panel: Same for 100\,km rock planetesimals.}
\label{1aep}
\end{figure}

\begin{figure}
\centerline{\includegraphics[width=14cm]{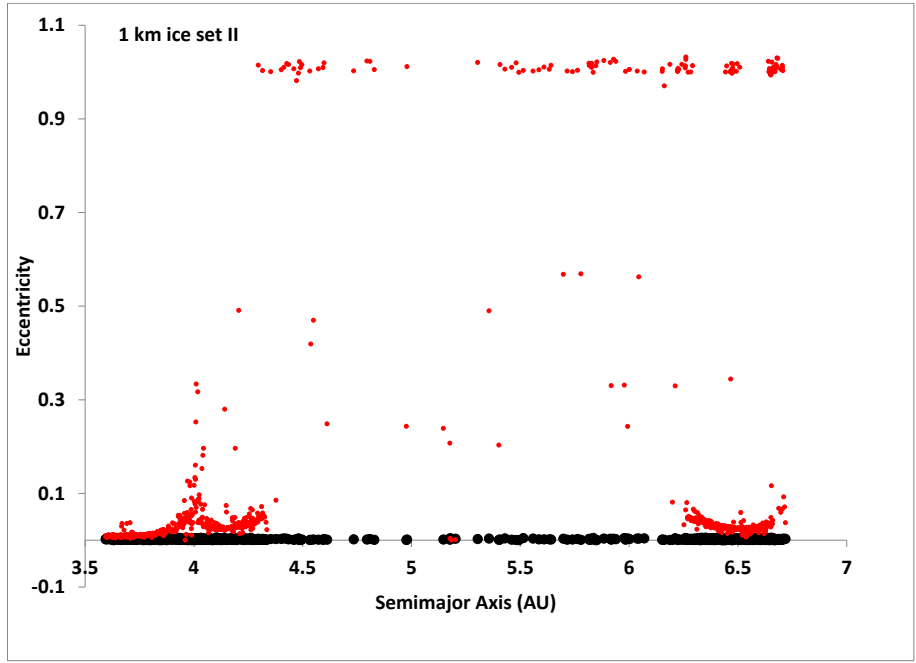}}
\centerline{\includegraphics[width=14cm]{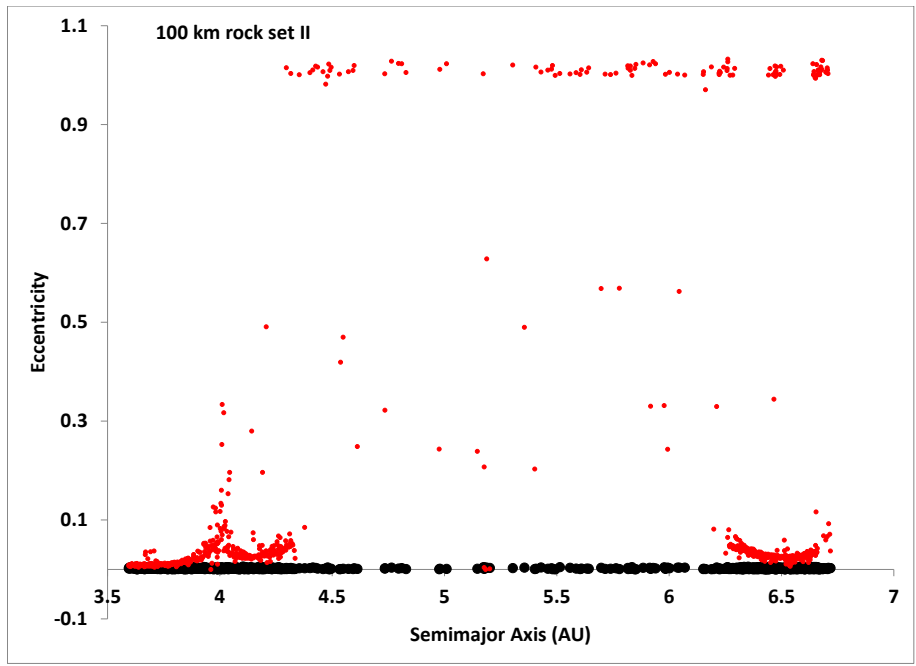}}
\caption{Same as Fig.\,\ref{1aep} for planetesimals of set II.}
\label{1aeIIp}
\end{figure}

The regions from where the accreted planetesimals originate are of interest because they enable us to develop an understanding of the connection between the chemical composition of the planetesimals and the rate of their capture. In addition, it allows us to determine the fate of those objects that are not accreted or are scattered to orbits with higher eccentricities.  Figure \ref{1aep} shows the initial (black dots) and final (red dots) eccentricities as a function of initial semimajor axis for 1\,km ice and 100\,km rock planetesimals of set I that are not accreted after $3\times 10^6$ years.  Most of the planetesimals between 4.7 and 5.5\,AU are accreted so that this region is much sparser than the regions farther from Jupiter.  Planetesimals that originated in this region and are not accreted are excited into more eccentric orbits or are given hyperbolic trajectories and are not shown.  Farther out on either side of Jupiter are planetesimals that have been excited into orbits with higher eccentricity, while planetesimals that originated more than around 1\,AU one either side of the protoplanet have their orbits essentially unchanged.  The result for case II planetesimals is shown in Figure \ref{1aeIIp}.  Once again, we see that the behavior of the 1\,km and 100\,km planetesimals is almost identical.  Mixed (ice + rock) planetesimals also behave in a very similar way.  This again shows that the protoplanet essentially sweeps up the region within approximately 1\,AU on either side of its orbit regardless of the size and composition of the planetesimals.

\section{Effect of Nebular Gas}
In all of the preceding computations we neglected the effect of nebular gas, however this is an important effect and has been the subject of a number of studies. \cite{zhou2007} studied the motion of planetesimals under the combined influence of the aerodynamic drag of the nebular gas and the tidal interaction with the disk.  They found that because of these effects the planetesimals generally do not pass through the mean motion resonances of the protoplanetary core, and the accretion rate slows as a result.  Further growth of the protoplanet causes the resonances to overlap, exciting the protoplanet eccentricities.  Gas drag then causes the planetesimals to collide with the protoplanet, causing a surge in the accretion rate.
	
\cite{shiraishi2008} developed a semi-analytical formula for the planetesimal accretion rate based on the competition between the growth of the protoplanet's Hill's sphere and the dampening of planetesimals eccentricities due to gas drag.  \cite{tanigawa2010} performed more detailed calculations of the effect of gas drag on the details of planetesimal capture.  More recent studies of the effect of gap formation in the nebular gas by the protoplanet have been presented in \cite{shibata2019}.  All of these studies show the importance of including nebular gas drag in the computation of planetesimal accretion.
	
However, the precise magnitude of the effect is dependent on the details of the density and motion of the nebular gas as a function of time.  While it is beyond the scope of this work to develop a detailed model for nebular gas motion, it is important to understand how the presence of nebular gas affects our results.  To this end, we have made some runs including nebular gas to see how the planetesimal behavior changes as a result.  We assume the following simple model for the gas disk:  The total mass of the disk is that of a minimum mass solar nebula (MMSN), 0.01 $M_{\odot}$.  Following \cite{hayashi1981} we assume that the gas surface density varies like $\sigma_g\sim r^{-3/2}$.  The temperature is assumed to vary like $T\sim r^{-3/4}$, corresponding to a slightly flared disk \citep{armitage07} so that the disk scale height varies as $H\sim r^{9/8}$.  As a result the midplane gas density varies like $\rho_g\sim r^{-21/8}$ and the gas pressure as $P\sim r^{=27/8}$.  This parameterization allows us to compute gas speed and the resultant drag force on the planetesimals \citep[see, e.g.][]{zhou2007}.  For the basic model we take $\rho_g=8.4\times 10^{-10}$\,\gcc and $T=280$\,K at 1\,AU.  We also assume these parameters remain constant in time.      
	
\begin{figure}
	\centerline{\includegraphics[width=18cm]{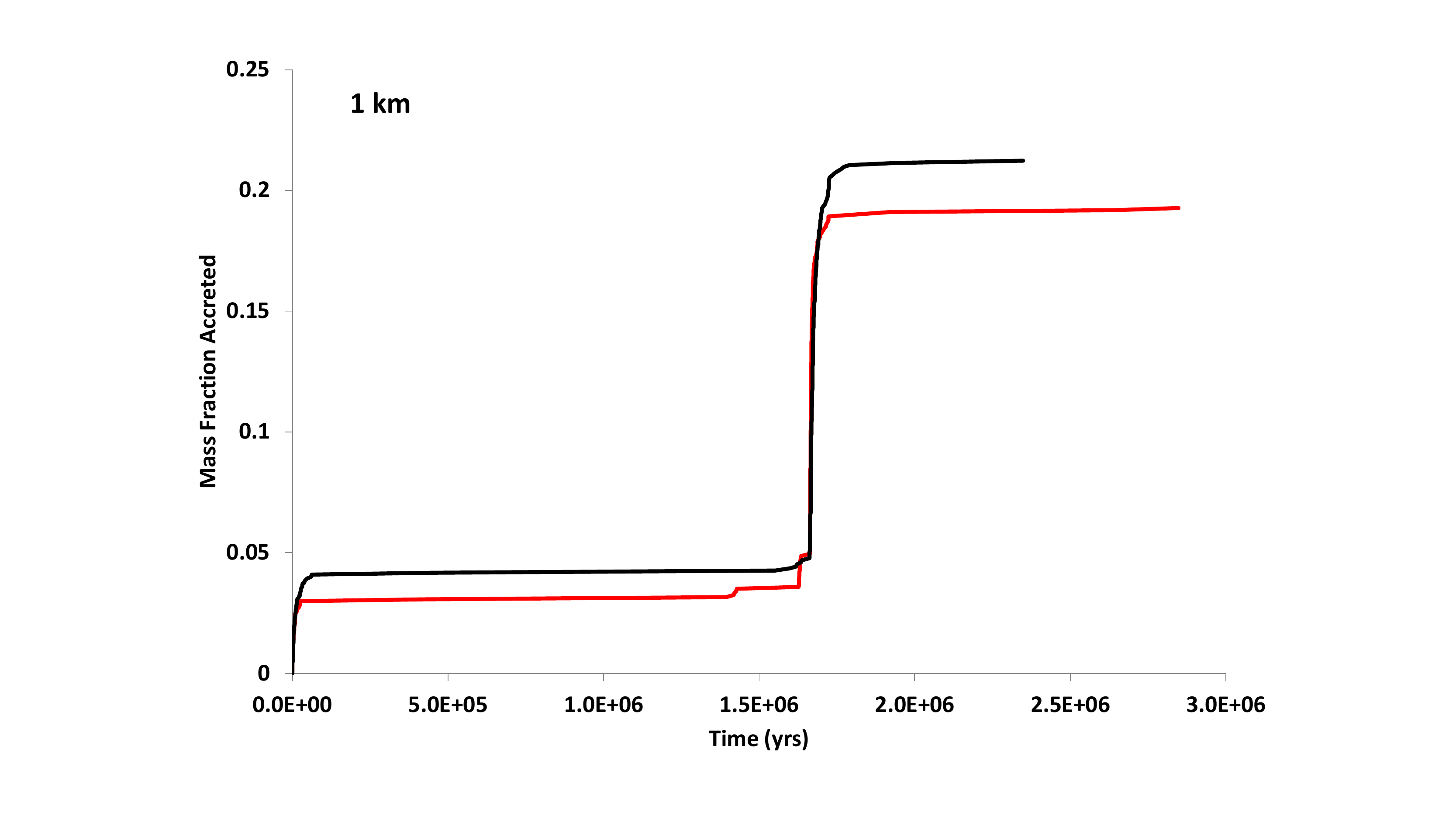}}
	\centerline{\includegraphics[width=18cm]{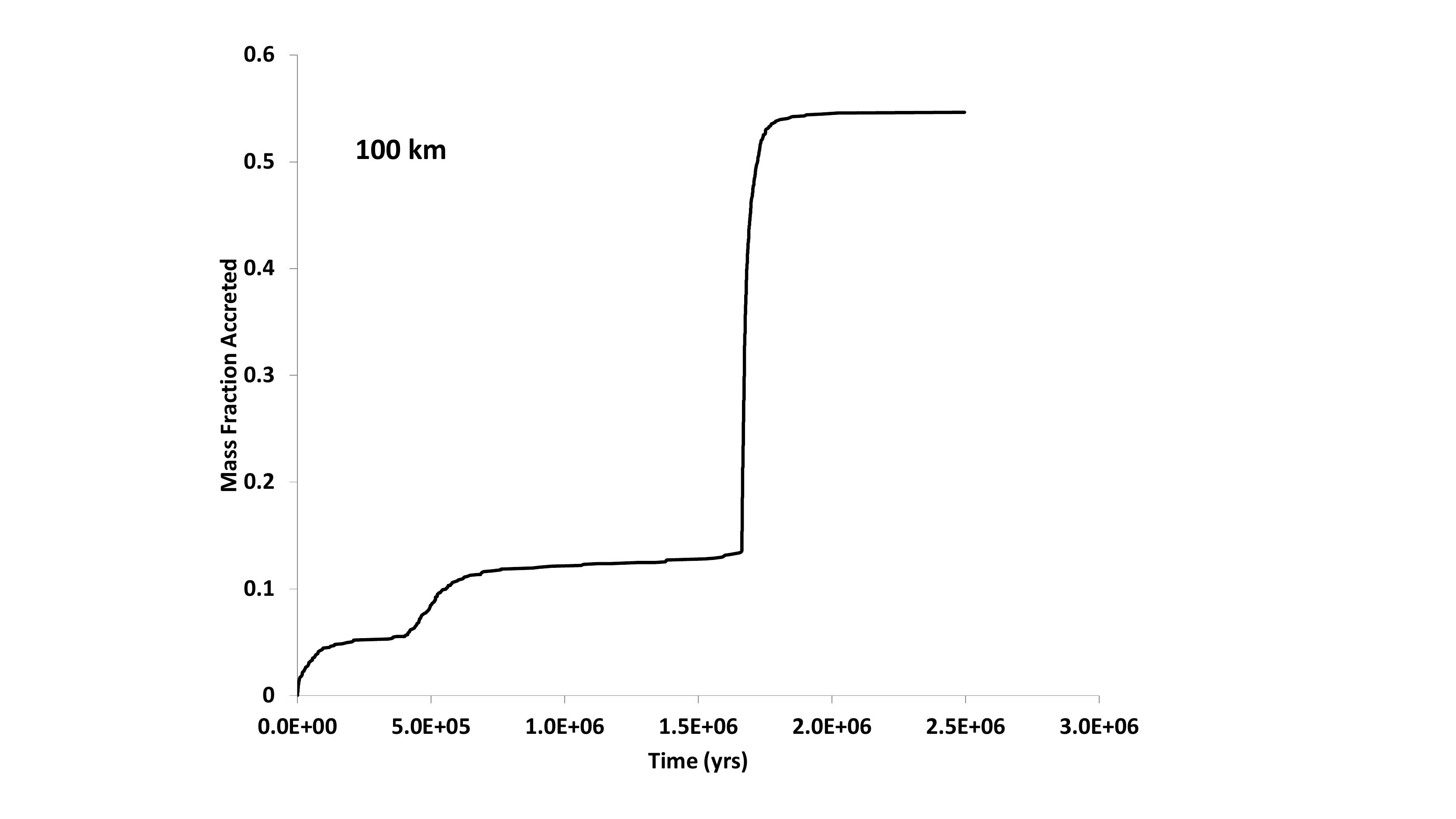}}
	\caption{Fraction of planetesimals captured as a function of time when nebular gas drag is included.  Upper panel: 1\,km planetesimals of ice (red) and rock (black).  Lower panel: 100\,km rock planetesimals.}
	\label{maccN}
\end{figure}

Figure\,\ref{maccN} shows the fraction of planetesimals accreted as a function of time, for the cases illustrated in figure\,\ref{maccp}, but with nebular gas drag included.  The upper panel shows the case for 1\,km planetesimals.  Both ice (red) and rock (black) planetesimals are strongly affected by the gas drag, and only 19\% of the ice and 21\% of the rock planetesimals are accreted by the protoplanet.  This decrease in the capture rate is due to the fact that those planetesimals with orbits inside the orbit of Jupiter that are not captured quickly, drift sunward due to the gas drag and out of the feeding zone of the protoplanet.  Those planetesimals with orbits outside Jupiter's orbit drift closer to the protoplanet, but, as noted by \cite{zhou2007}, do not spend significant time at the mean motion resonances of the protoplanetary core, and are not captured.  This can be seen in the long flat region of the curve between the initial growth phase near time zero and the rapid spurt at the time of runaway growth.  This can also be seen, to a smaller extent, in the lower panel for the case of 100\,km rock planetesimals.  Although these are much less affected by the gas drag, the growth curve is still quite flat between about $6\times 10^5$\,y and the time of runaway growth.
		
\begin{figure}
	\centerline{\includegraphics[width=20cm]{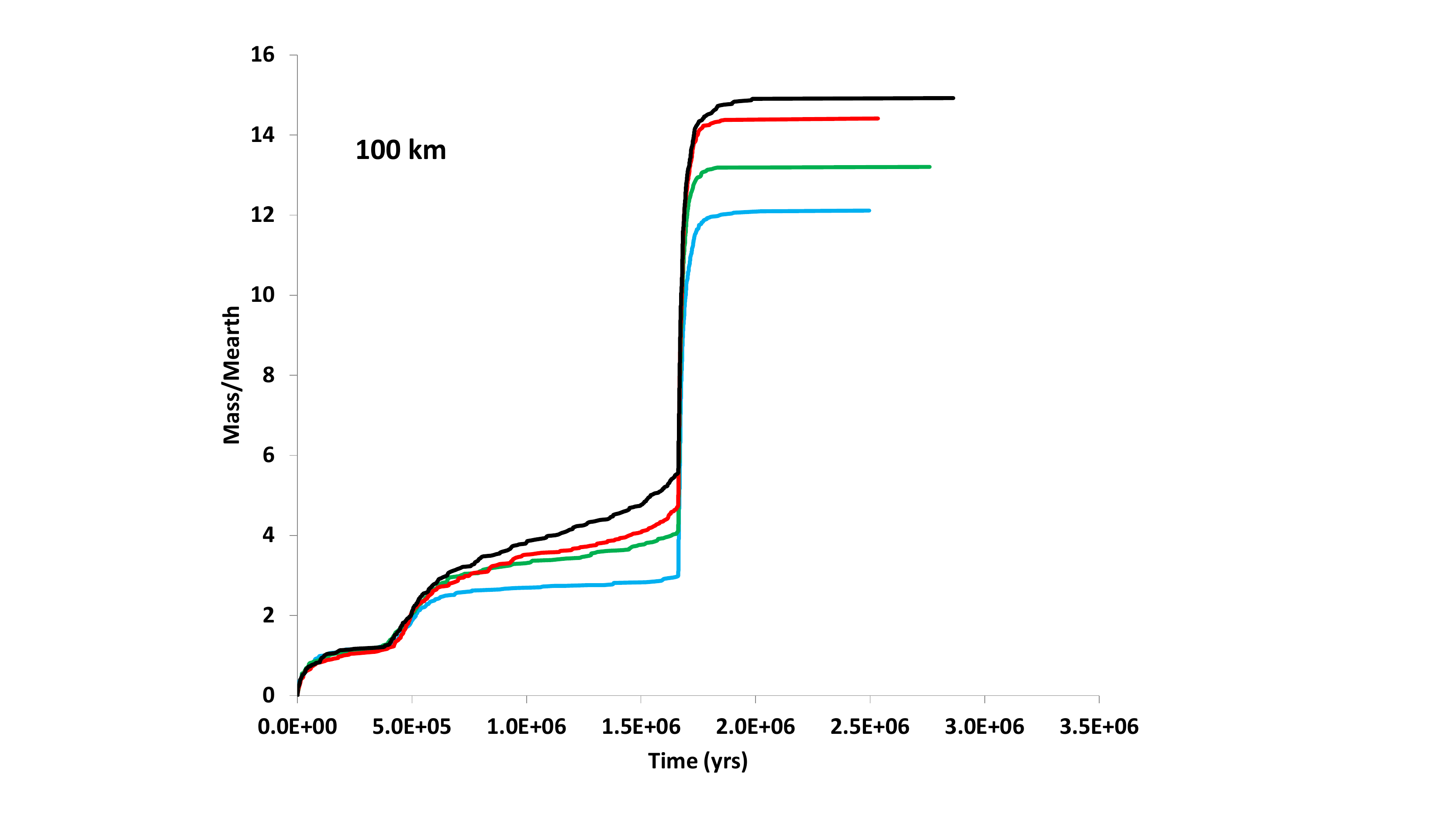}}
	\caption{Mass (in Earth masses) of 100\,km rock planetesimals captured as a function of time for different values of the nebular gas density.  Shown are runs for a midplane density of $\rho_g=8.4\times 10^{-10}$\,\gcc at 1\,AU (blue), $2.1\times 10^{-10}$\,\gcc, (green), $1.05\times 10^{-10}$\,\gcc, and 0 (black).}
	\label{maccNv}
\end{figure}

The nebular model we have investigated assumes that the gas density remains constant in time.  In order to get a feeling for the effect of lowering the density as the gas dissipates, we have run two additional cases for 100\,km rock planetesimals; one where the gas density at 1\,AU is taken to be $2.1\times 10^{-10}$\,\gcc, and one where it is taken to be $1.05\times 10^{-10}$\,\gcc.  Figure\,\ref{maccNv} shows the mass of accreted planetesimals for all four cases we explored; $\rho_{g}(1\rm{AU})=8.4\times 10^{-10}$\,\gcc (blue curve) to $2.1\times 10^{-10}$\,\gcc (green curve), $1.05\times 10^{-10}$\,\gcc (red curve), and 0 (black curve).

\section{Summary and Conclusions}
We have computed the motion of planetesimals in the presence of the Sun and a growing Jupiter in order to calculate a more accurate mass accretion rate for the growing protoplanet.  In addition to the gravitational forces of the two large bodies, the effect of gas drag and ablation of the proto-Jovian envelope was included. Because the planetesimals encounter the envelope with relatively high speeds, especially in the later stages of protoplanet growth, we find that, based on the assumed internal strength of the materials considered, the ram pressure is high enough to cause planetesimals to break up, contrary to the findings of \cite{inaba03b}.  We found that during the first $5\times 10^5$\,yr the mass accretion rate was lower than that assumed by \cite{lozovsky17}.  This results in a significantly smaller core than they assumed.  As a result, the mass of the gaseous envelope will probably be smaller than what we have assumed in this simulation.  Certainly the protoplanet evolution and the planetesimal capture have to be computed self-consistently for a proper assessment of the evolutionary history of the planet.

In spite the lack of self-consistency in our simulation, we believe that the rapid accretion of solids at the time of rapid gas accretion is real and important feature of the CA scenario.  This means that, regardless of when the collapse occurs, the rapid infall of gas, together with the spike in the planetesimal accretion rate would allow for a much larger fraction of the accreted material to remain in the envelope.  In addition, after collapse, the accretion rate obtained from our new model is significantly higher than that assumed for model A.  For the assumed surface density of $\sigma=6$\,\gcms\ we find that the total mass accreted is 14.9\,$M_{\oplus}$ for set I and 13.5\,$M_{\oplus}$ for set II, while model A has a total accreted heavy element mass of only 10.6\,$M_{\oplus}$.  Again, one would expect that much of this late-accreted material would remain in the envelope, in agreement with the Juno gravity field measurements.  

It is important to note that in this study, the effect of Saturn was not included. It would not be implausible to assume that including the effect of Saturn may change some of the results. A second effect is the drag of the nebular gas itself.  As noted above, nebular gas drag will be more important for smaller planetesimals, and particularly for pebble accretion.  Thus our results are far from complete.  However, the details of the motion of the gas in the vicinity of the protoplanet are complex and beyond the scope of this work.  These effects will be the subject of future investigations.

\section*{Acknowledgments}
We would like to thank Drs. Matt Holman and William Newman for helpful discussions.  We would also like to thank the anonymous referee for critically reading our manuscript and useful comments that have improved our paper. MP acknowledges the support of ISF grant 566/17. NH would like to acknowledge support from NASA grant 80NSSC18K0518.  PB acknowledges support from NASA grant NNX14AG92G

\appendix
\section*{Appendix-Details of Ablation Calculation}
Here we summarize the calculation of the planetesimal's interaction with the gas.  Further details are given in \cite{podolak88}.

If $a$ is the planetesimal radius and $\lambda$ is the mean free path of a gas molecule, the Knudsen number is given by
\begin{equation}\label{knud}
Kn=\frac{\lambda}{a}
\end{equation}
The viscosity of the gas will be given by
$$\eta=2.38\times 10^{-6} T_g^{2/3}\ \ \ \ {\rm g\,cm^{-1}\,s^{-1}}$$
where $T_g$ is the gas temperature. The Reynolds number will then be
\begin{equation}\label{reyn}
Re=\frac{\rho_g v_r a}{\eta}
\end{equation}
where $v_r$ is the relative velocity of the planetesimal with respect to the protoplanet.

The speed of sound in the gas is
$$c_s=\sqrt{\frac{1.4kN_AT_g}{\mu_g}}$$
where $N_A$ is Avogadro's number, $k$ is Boltzmann's constant, $\mu_g$ is the mean molecular weight of the gas.  The Mach number is then 
\begin{equation}
Ma=\frac{v_r}{c_s}
\end{equation}
The drag force, $\vec{F}_d$, is computed as follows:

If $Re<1$
$$\psi=1+Kn\left[1.249+0.42e^{-0.87/Kn}\right]$$
The $i$th component of the drag force is then
\begin{equation}\label{lam}
F_{di}=\frac{6\pi a\eta }{\psi}v_{ri}
\end{equation}
where $v_{ri}$ is the $i$th component of the relative velocity vector.

If $Ma\leq 1,\ Re>1$
\begin{equation}\label{nonlam}
F_{di}=D\pi a^2\rho_g v_r v_{ri}
\end{equation}
Here $D$ is a drag coefficient which can be approximated as follows:
$$D=\frac{6}{\sqrt{Re}}\ \ \ \ \ 1\leq Re\leq 10^3$$
\begin{equation}\label{d}
\ \ \ \ \ =0.19 \ \ \ \ \ \ \ \ 10^3<Re\leq 10^5
\end{equation}
$$\ \ =0.15 \ \ \ \ \ \ \ Re>10^5$$

If  $Ma>1$
$$D=1.1-\frac{\log (Re)}{6}\ \ \ \ \ 1\leq Re\leq 10^3$$
\begin{equation}\label{ssonic}
\ =0.5 \ \ \ \ \ \ \ \ \ \ \ \ \ \ \ Re>10^3
\end{equation}

As a result of gas drag the planetesimal will be heated and will lose material.  The rate of mass loss will depend temperature, $T_p$ of the planetesimal.  The rate at which energy is transfered to the planetesimal by gas drag is 
$$E_d=\frac{\alpha}{2}\rho_gv^3_r\ \ \ \ {\rm erg\,cm^{-2}\,s^{-1}}$$
The cross-sectional area of the planetesimal is $\pi a^2$, so the energy delivered per unit area of surface is $0.25E_d$.  We take $\alpha=D/2$. In addition the planetesimal is heated over its entire surface by radiation from the surrounding gas at the rate 
$$E_T=\sigma T_g^4\ \ \ \ {\rm erg\,cm^{-2}\,s^{-1}}$$ where $\sigma$ is the Stefan-Boltzmann constant.  The total rate of heat input is therefore 
\begin{equation}\label{heat}
E=\frac{E_d}{4}+E_T
\end{equation}

This heat input is balanced by reradiation to space and by evaporation.  The rate of evaporation is given by
$$\xi=P_{vap}\sqrt{\frac{\pi m_p}{8kT_p}}$$
Here $m_p$ is the mass of a planetesimal molecule and $k$ is Boltzmann's constant.  $P_{vap}$ is the vapor pressure and it is given by 
$$P_{vap}=P_0e^{-A/T_p}$$ It too depends on the composition of the planetesimal via $P_0$ and $A$.  When the heating equals the cooling
$$\frac{E_d}{4}+E_T=\frac{\alpha}{8}\rho_gv^3_r+\sigma T_g^4=E_0P_{vap}\sqrt{\frac{\pi m_p}{8kT_p}}+\sigma T_p^4$$

The rate of mass loss is given by
$$\frac{dM_p}{dt}=4\pi a^2\xi=4\pi a^2\rho_p\frac{da}{dt}$$ we get
\begin{equation}\label{dadt}
\frac{da}{dt}=\frac{P_{vap}}{\rho_p}\sqrt{\frac{\pi m_p}{8kT_p}}
\end{equation}

\begin{table}[h] \centering
	\begin{tabular}{|c|c|c|c|c|}\hline
		Parameter &  Ice    &   Rock  & Iron & Mixed \\ \hline
		$m_p$   & $2.99\times 10^{-23}$ & $8.31\times 10^{-23}$ & $9.3\times 10^{-23}$ & $2.99\times 10^{-23}$  \\
		$\rho_p$   & 1.0 & 3.4 & 7.8 & 1.0 \\
		$E_0$   & $2.8\times 10^{10}$ & $8.08\times 10^{10}$ & $8.26\times 10^{10}$  &  $2.8\times 10^{10}$  \\
		$P_0$   & $3.891\times 10^{11}$ & $1.50\times 10^{13}$ & $3.221\times 10^{12}$ & $3.891\times 10^{11}$ \\
		$A$ & $-2.1042\times 10^3$ & $-2.4605\times 10^4$ & $-2.0014\times 10^{4}$ & $-2.1042\times 10^3$ \\
		$P_{crit}$ & $10^6$  & $10^8$ & $10^8$ & $10^6$ \\
		$T_{crit}$  & 648 & 4000 & 4000 & 648 \\ \hline
		\hline
	\end{tabular}
	\caption{Physical parameters for different compositions.  All values are in cgs units.}
\end{table}

Eq.\,(\ref{dadt}) holds as long as $T_p\leq T_{crit}$.  For $T_p>T_{crit}$ the material behaves like a gas, and will simply flow off the grain.  The rate of mass loss will then be given by the rate at which the material can be heated to $T_{crit}$.  
\begin{equation}\label{dadtcrit}
\frac{da}{dt}=\frac{E-\sigma T_{crit}^4}{E_0\rho_p}
\end{equation}
where $E$ is given by Eq.\,(\ref{heat}).  The values for $T_{crit}$ and other relevant parameters are given in the table.  When $a\leq 0.2a_{initial}$ we can assume that the planetesimal is completely evaporated and the integration is stopped.

Another process that has to be considered is breakup due to the ram pressure exerted on the planetesimal by the gas.  If the dynamic pressure, $P_d$ is greater than the material strength, and the self gravity of the planetesimal is not enough to hold the planetesimal together, the planetesimal will fragment.  The dynamic pressure is given by
\begin{equation}
P_d=\frac{v_r^2\rho_g}{2}
\end{equation}
and the dynamic radius, $R_d$, at which self gravity is sufficiently strong to hold the planetesimal together is
\begin{equation}
R_d=\frac{\sqrt{5.96\times 10^6P_d}}{\rho_p}
\end{equation}
The criterion for fragmentation is then $$P_d > P_{crit}$$ and $$R_d > a$$

\newpage

\begin{figure}
	\centerline{\includegraphics[width=18cm]{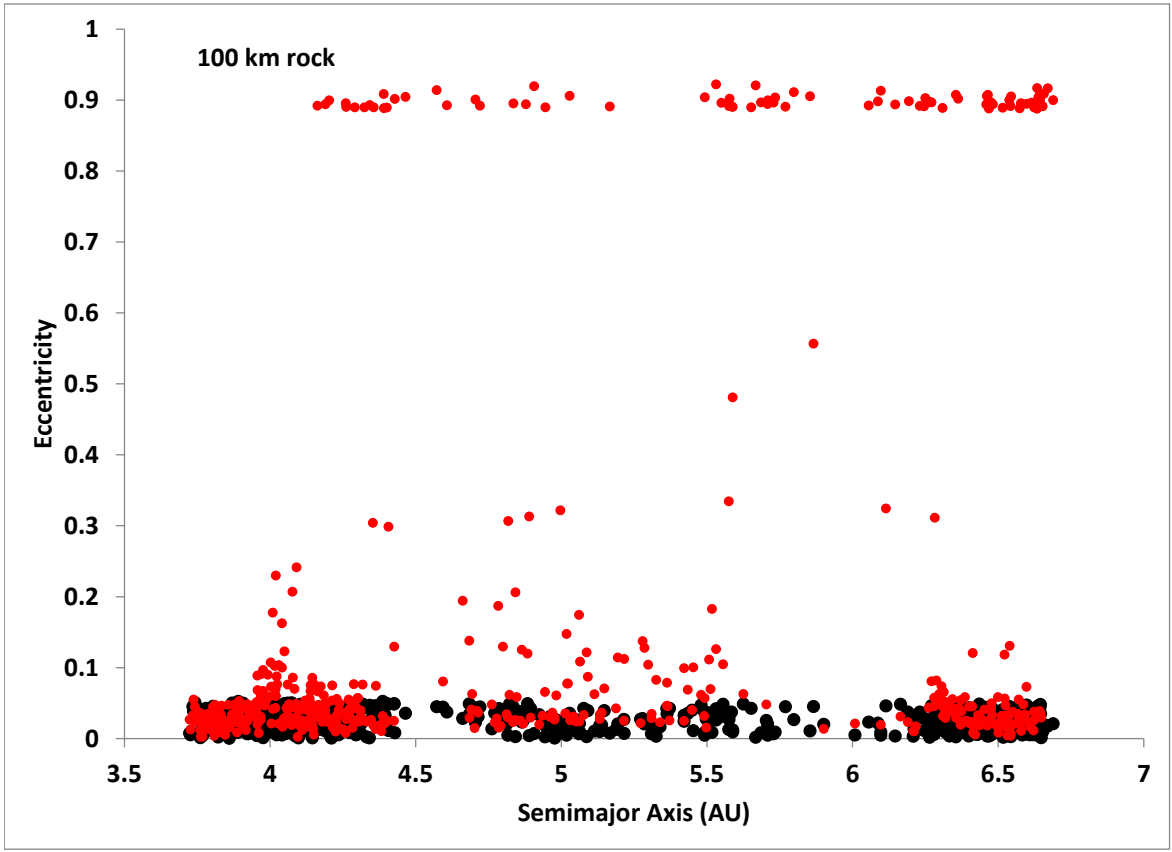}}
	\caption{Initial (black) and final (red) eccentricities of 100\,km planetesimals that are not captured as a function of their initial semimajor axis in the presence of nebular gas.  This should be compared with the lower panel of Fig.\,\ref{1aep}   .}
	\label{1aepN}
\end{figure}

\end{document}